\documentclass[a4paper,11pt]{article}
\pdfoutput=1 

\usepackage{jcappub} 

\usepackage[T1]{fontenc} 
\usepackage{verbatim}
\usepackage[utf8]{inputenc}
\usepackage[american]{babel}
\usepackage{graphicx}
\usepackage{epsfig}
\usepackage{adjustbox}
\usepackage{booktabs}
\usepackage{multirow}
\usepackage{dcolumn}
\usepackage{amsmath}
\usepackage{amsfonts}
\usepackage{mathtools}
\usepackage{amssymb}
\usepackage{subcaption} 
\usepackage{caption}
\usepackage{epstopdf}
\usepackage[font=small]{caption}
\usepackage{bm}
\usepackage{mathrsfs}
\usepackage{version}
\usepackage{enumerate}
\usepackage{braket}
\usepackage{enumitem}
\usepackage{transparent}
\usepackage{pifont}
\usepackage{colortbl}
\usepackage{rotating}
\usepackage{adjustbox}
\usepackage{cancel}
\usepackage{tabularx}
\usepackage{soul}
\usepackage[table]{xcolor}
\usepackage{siunitx}
\usepackage{float} 
\usepackage{cite} 

\usepackage{hyperref}
\hypersetup{colorlinks=true,
linkcolor=navyblue,citecolor=navyblue,urlcolor=navyblue,pdfencoding=auto}

\definecolor{navyblue}{rgb}{0.0, 0.0, 0.5}
\definecolor{royalblue}{rgb}{0.25, 0.41, 0.88}
\definecolor{cadmiumgreen}{rgb}{0.0, 0.42, 0.24}
\definecolor{blue-violet}{rgb}{0.54, 0.17, 0.89}
\definecolor{darkviolet}{rgb}{0.58, 0.0, 0.83}
\definecolor{orange(colorwheel)}{rgb}{1.0, 0.5, 0.0}

\definecolor{magenta(process)}{rgb}{1.0, 0.0, 0.56}

\definecolor{darkspringgreen}{rgb}{0.09, 0.45, 0.27}

\definecolor{royalblue(web)}{rgb}{0.25, 0.41, 0.88}

\definecolor{cadmiumorange}{rgb}{0.93, 0.53, 0.18}

\definecolor{heliotrope}{rgb}{0.87, 0.45, 1.0}

\makeatletter
\renewcommand*{\@textcolor}[3]{%
\protect\leavevmode
\begingroup
\color#1{#2}#3%
\endgroup
}
\makeatother

\newcommand\horsp{\rule[-1.5mm]{0mm}{4.125mm}}
\newcommand\morehorsp{\rule[-2.25mm]{0mm}{6mm}}

\renewcommand\({\left(}
\renewcommand\){\right)}
\renewcommand\[{\left[}

\makeatletter
\let\save@mathaccent\mathaccent
\newcommand*\if@single[3]{%
\setbox0\hbox{${\mathaccent"0362{#1}}^H$}%
\setbox2\hbox{${\mathaccent"0362{\kern0pt#1}}^H$}%
\ifdim\ht0=\ht2 #3\else #2\fi
}
\newcommand*\rel@kern[1]{\kern#1\dimexpr\macc@kerna}
\newcommand*\widebar[1]{\@ifnextchar^{{\wide@bar{#1}{0}}}{\wide@bar{#1}{1}}}
\newcommand*\wide@bar[2]{\if@single{#1}{\wide@bar@{#1}{#2}{1}}{\wide@bar@{#1}{#2}{2}}}
\newcommand*\wide@bar@[3]{%
\begingroup
\def\mathaccent##1##2{%
\let\mathaccent\save@mathaccent
\if#32 \let\macc@nucleus\first@char \fi
\setbox\z@\hbox{$\macc@style{\macc@nucleus}_{}$}%
\setbox\tw@\hbox{$\macc@style{\macc@nucleus}{}_{}$}%
\dimen@\wd\tw@
\advance\dimen@-\wd\z@
\divide\dimen@ 3
\@tempdima\wd\tw@
\advance\@tempdima-\scriptspace
\divide\@tempdima 10
\advance\dimen@-\@tempdima
\ifdim\dimen@>\z@ \dimen@0pt\fi
\rel@kern{0.6}\kern-\dimen@
\if#31
\overline{\rel@kern{-0.6}\kern\dimen@\macc@nucleus\rel@kern{0.4}\kern\dimen@}%
\advance\dimen@0.4\dimexpr\macc@kerna
\let\final@kern#2%
\ifdim\dimen@<\z@ \let\final@kern1\fi
\if\final@kern1 \kern-\dimen@\fi
\else
\overline{\rel@kern{-0.6}\kern\dimen@#1}%
\fi
}%
\macc@depth\@ne
\let\math@bgroup\@empty \let\math@egroup\macc@set@skewchar
\mathsurround\z@ \frozen@everymath{\mathgroup\macc@group\relax}%
\macc@set@skewchar\relax
\let\mathaccentV\macc@nested@a
\if#31
\macc@nested@a\relax111{#1}%
\else
\def\gobble@till@marker##1\endmarker{}%
\futurelet\first@char\gobble@till@marker#1\endmarker
\ifcat\noexpand\first@char A\else
\def\first@char{}%
\fi
\macc@nested@a\relax111{\first@char}%
\fi
\endgroup
}
\makeatother

\newcommand\ee{\end{equation}}
\newcommand\be{\begin{equation}}
\newcommand\eea{\end{eqnarray}}
\newcommand\bea{\begin{eqnarray}}
\newcommand{\bsp}{\begin{split}}
\newcommand{\esp}{\end{split}}
\newcommand{\bit}{\begin{itemize}[leftmargin=*]}
\newcommand{\eit}{\end{itemize}}
\newcommand{\ben}{\begin{enumerate}[leftmargin=*]}
\newcommand{\een}{\end{enumerate}}

\newcommand{\ie}{
i.e.
~}

\newcommand\eq[1]{Eq.~\eqref{eq:#1}}

\renewcommand{\vec}{\bm}

\newcommand{\Bmpc}{B_{1\,\rm Mpc}}

\title{The impact of primordial magnetic fields on future CMB bounds on inflationary gravitational waves}

\author[a]{Fabrizio Renzi,}
\author[b]{Giovanni Cabass,}
\author[c]{Eleonora Di Valentino,}
\author[a]{Alessandro Melchiorri,}
\author[d,e,f]{Luca Pagano}

\affiliation[a]{Physics Department and INFN, Universit\`a di Roma ``La Sapienza'', Ple Aldo Moro 2, 00185, Rome, Italy}
\affiliation[b]{Max-Planck-Institut f\"{u}r Astrophysik, 
Karl-Schwarzschild-Str. 1, 85741 Garching, Germany}
\affiliation[c]{Jodrell Bank Center for Astrophysics, School of Physics and Astronomy, University of Manchester, Oxford Road, Manchester, M13 9PL, UK}
\affiliation[d]{Institut d'Astrophysique Spatiale, CNRS, Univ. Paris-Sud, Universit\'{e} Paris-Saclay, B\^{a}t. 121, 91405 Orsay cedex, France}
\affiliation[e]{Institut d'Astrophysique de Paris, CNRS, 98 bis Boulevard Arago, F-75014, Paris, France}
\affiliation[f]{LERMA, Sorbonne Universit\'{e}, Observatoire de Paris, Universit\'{e} PSL, \'{E}cole normale sup\'{e}rieure, CNRS, Paris, France}

\emailAdd{fabrizio.renzi@uniroma1.it}
\emailAdd{gcabass@mpa-garching.mpg.de}
\emailAdd{eleonora.divalentino@manchester.ac.uk}
\emailAdd{alessandro.melchiorri@roma1.infn.it}
\emailAdd{lpagano@ias.u-psud.fr}

\abstract{\noindent We discuss whether an unaccounted contribution to the Cosmic Microwave Background polarization $B$-mode 
by primordial magnetic fields (PMFs) can bias future constraints on inflationary gravitational waves. 
As a case-study, we consider a scale-invariant PMF spectrum with amplitude of $\approx{1}\,{\mathrm{nG}}$ on $1\,\mathrm{Mpc}$ scales, 
compatible with current cosmological bounds. 
We find a degeneracy in the $B$-mode spectra between PMFs and inflationary gravitational waves. 
If PMFs of this amplitude are not accounted for, future CMB experiments could claim a false detection of a tensor-to-scalar ratio $r\approx 0.007$, 
close to the predictions of Starobinsky and $\alpha$-attractor models. 
The degeneracy can be broken if $B$-modes are measured also at multipoles $\ell\gtrsim 900$: 
more precisely experiments like CMB-S4 or CORE-M5 would be able to discriminate PMFs from primordial GWs at high statistical significance. 
Experiments like LiteBIRD or PIXIE will not be able to break the degeneracy and will need complementary bounds coming, for example, 
from measurements of anisotropies in the Faraday rotation angle of CMB polarization. This reinforces the importance of future experimental constraints on PMFs. }

\begin{document}
\maketitle
\flushbottom


\section{Introduction}
\label{sec:intro}

\noindent One of the main goals of modern cosmology is the detection of Cosmic Microwave Background polarization $B$-modes produced by 
vacuum fluctuations of the metric during inflation. Their detection would provide a ``smoking gun'' for the inflationary paradigm and 
give hints towards the quantum nature of gravity. 

In the past years the experimental bounds on the primordial $B$-mode component (parameterized by the tensor-to-scalar ratio $r$) have improved significantly. 
Indeed, since the constraints from the BICEP experiment of $r<0.72$ at $95 \%$ C.L. \cite{Chiang:2009xsa} in 2010, 
the recent combined analysis of BICEP2, Keck Array IV and Planck $B$-mode measurements now provide $r<0.07$ at $95 \%$ C.L. \cite{Array:2015xqh}, 
showing an improvement by nearly one order of magnitude in about $\sim 7$ years (see also \cite{Cabass:2015jwe}). 
In the next years a further improvement by one order of magnitude, reaching a sensitivity in the range of $r\sim 10^{-2}\,\text{-}\, 10^{-3}$, 
is expected by several ongoing experiments such as BICEP3 and the Keck Array \cite{Grayson:2016smb}, 
CLASS \cite{Essinger-Hileman:2014pja}, Advanced ACTPol \cite{Henderson:2015nzj}, and SPT-3G \cite{Benson:2014qhw}. 
Future experiments as the LiteBIRD \cite{Suzuki:2018cuy} 
and CORE-M5 \cite{DiValentino:2016foa,Finelli:2016cyd,Delabrouille:2017rct} satellite missions and the CMB-S4 ground based telescope 
\cite{Abazajian:2016yjj} are expected to reach a sensitivity of $\delta r \sim 0.0001$, 
closing in on the prediction $r\approx\num{d-3}$ of the Starobinsky model \cite{Starobinsky:1980te}. 

It is however important to investigate if other mechanisms 
could generate a $B$-mode polarization signal that could lead to a wrong claim for a detection of vacuum fluctuations of the metric. 
For example, foregrounds as galactic dust are obviously an issue \cite{Akrami:2018wkt}. 
Topological defects (see e.g. \cite{Lizarraga:2014eaa}) can also produce $B$-modes (even from vector perturbations \cite{Moss:2014cra}). 
Finally, GWs can also be sourced during inflation in presence of anisotropic stress generated by quantum fluctuations of other fields, 
even if their energy density is much smaller than that of the dominant inflaton field (see e.g. \cite{Pajer:2013fsa,Namba:2015gja}). 

The common attribute to all these sources is the fact that they have some additional signature that allows to 
disentangle them from $B$-modes generated from vacuum fluctuations of the metric, 
be it the frequency dependence for galactic foregrounds (see e.g. \cite{Errard:2015cxa, Remazeilles:2017szm}), 
or the shape of the angular power spectrum for $B$-modes from topological defects or sourced tensor fluctuations during inflation.\footnote{
Moreover, sourced tensor fluctuations can be chiral \cite{Lue:1998mq,Saito:2007kt,Gluscevic:2010vv,Gerbino:2016mqb,Thorne:2017jft} 
and highly non-Gaussian \cite{Agrawal:2017awz,Agrawal:2018mrg}. }

$B$-mode polarization can also be produced by a primordial magnetic field (PMF) 
(see e.g. \cite{Shaw:2009nf,Bonvin:2014xia,Giovannini:2017rbc}). 
In the presence of PMFs, passive tensor and compensated vector modes give $B$-modes with angular spectra 
that are very similar in shape to those produced by primordial gravitational waves (GWs)\footnote{For simplicity, in the rest of the paper 
we will refer to vacuum fluctuations of the metric during inflation as primordial gravitational waves.} 
and lensing (see e.g. \cite{Finelli:2008xh,Paoletti:2008ck,Shaw:2009nf,Zucca:2016iur}). 
Future CMB experiments like CMB-S4 will be extremely sensitive to PMFs, 
improving current constraints on the corresponding $B$-mode amplitude by nearly two orders of magnitude \cite{Sutton:2017jgr}. 

It is then important to investigate how well future CMB experiments could discriminate between inflationary 
GWs and PMFs in the generation of CMB polarization $B$-modes: 
\emph{these experiments would be able to claim a detection of the quantum nature of tensor perturbations of the metric 
only if a potential contribution from PMFs can be identified and subtracted.} 

How can one distinguish between the two scenarios? 
One key difference between the two is that while GWs affect the CMB anisotropies on large angular scales, 
primordial magnetic fields affect also smaller angular scales through compensated perturbations 
(see e.g. \cite{Finelli:2008xh,Shaw:2009nf,Zucca:2016iur,Pogosian:2018vfr}). 
As we show later in this paper, CMB experiments sensitive mainly to large angular scales as the proposed PIXIE and LiteBIRD missions 
are essentially unable to discriminate between PMFs and primordial GWs {at the level of $r\approx\num{d-3}$}. 
Then, a clean and reliable detection of $B$-modes generated by the inflationary tensor fluctuations 
can be obtained only by considering also smaller scales, as planned by the CMB-S4 experiments, 
or by a satellite with improved angular resolution as the recent CORE-M5 proposal. 

A second difference between these two scenarios is the fact that PMFs 
can be constrained also by measuring the Faraday rotation (FR) of CMB polarization (see e.g. \cite{Pogosian:2018vfr} and references therein) 
considering maps at different frequencies and taking advantage of the fact that the frequency scaling of FR is $\sim\nu^{-2}$. 
Moreover, FR can be in principle measured either by considering the effect on CMB anisotropies angular spectra or by 
extracting the Faraday rotation angle through estimators that make use of the coupling between $E$- and $B$-modes 
induced by FR \cite{Pogosian:2011qv,De:2013dra,Ade:2015cao,Array:2017rlf}. 
It is therefore important to evaluate whether future realistic CMB experiments, considering their sensitivities, angular resolution and 
frequency and sky coverages, would be able to use FR to differentiate PMFs from inflationary GWs. 

Our paper is structured as follows: 
in Section \ref{sec:2} we review how primordial magnetic fields affect CMB temperature and polarization anisotropies and 
discuss in what region of parameter space they give a $B$-mode signal degenerate with that from primordial GWs. 
In Section \ref{sec:3} we review our forecast method while in Section \ref{sec:4} we present our results. 
In Section \ref{sec:5} we discuss the impact of Faraday rotation {on our forecasts}. We finally conclude in \mbox{Section \ref{sec:concl}. }

\section{Primordial magnetic fields}
\label{sec:2}

\noindent In this section we present some definitions useful for our analysis: we refer the interested reader to 
\cite{Finelli:2008xh,Paoletti:2008ck,Shaw:2009nf,Durrer:2013pga,Subramanian:2015lua,Grasso:2000wj} for a more detailed discussion of the subject.

\subsection{Definition of magnetic parameters}
\label{sec:general_definitions}

\noindent We consider a stochastic magnetic field $ B^i(\eta,\vec{x}) $ generated at a time $ \eta_B $ before the epoch of neutrino decoupling $ \eta_\nu $. 
We assume that PMFs are a statistically isotropic Gaussian field with no helicity (a review on the impact of helical field on CMB physics can be found in \cite{Subramanian:2015lua}). 
The exact form of the power spectrum $ P_B $ of the PMF strongly depends on the mechanism generating it. 
Following the current literature, we define $ P_B $ as a power law with a cut-off scale $ k_D $, i.e. 
\begin{equation}
\label{eq:PS_B}
P_B(k) =
\begin{cases}
\mathcal{A}\,k^{n_B} & \text{for $k < k_D\,\,,$} \\
0 & \text{otherwise}\,\,,
\end{cases}
\end{equation}
where a spectral index $n_B=-3$ denotes a scale-invariant spectrum 
and the cut-off scale $ k_D $ accounts for the damping of the magnetic field due to radiation viscosity on very small scales, 
where magnetic effects are suppressed by photon diffusion \cite{Mack:2001gc,Subramanian:1998fn}. 
From \eq{PS_B}, then, we define the magnetic field amplitude in terms of the parameter $ B_\lambda $, 
obtained by smoothing the magnetic energy density with a Gaussian filter over a comoving scale $ \lambda$ \cite{Shaw:2009nf}. 

In what follows we take $\lambda = 1\,\mathrm{Mpc} $.\footnote{It is common to take this value for $\lambda$ \cite{Ade:2015cva}: 
it corresponds to the size of a typical region at the time of last scattering that later collapses to form a galactic halo.} 
Moreover, we focus only on nearly scale-invariant spectra PMFs with $n_B = -2.9$, which we expect to be produced by 
inflationary magnetogenesis, where magnetic fields are generated on small scales and then stretched to cosmological scales by the accelerated expansion 
(see \cite{Kobayashi:2014sga,Kobayashi:2014zza,Green:2015fss} for recent analyses). 
We refer to \cite{Caprini:2001nb,Bonvin:2013tba,Camera:2013fva,Durrer:2013pga,Subramanian:2015lua,Ade:2015cva,Zucca:2016iur} 
for discussions about how values of $n_B$ larger than $-3$ can be generated, and what are the current constraints on PMFs with blue-tilted spectra. 

Depending on the generation epoch of the PMF three different class of magnetic perturbations can be distinguished: 
inflationary, passive and compensated. 
In this paper we focus only on passive and compensated mode which are sourced by every PMF independently of the magnetic generation history. 
Passive modes are generated before neutrino decoupling ($ \eta < \eta_\nu $) since, without neutrino free-streaming, 
there is no counterpart in the photon-baryon fluid able to compensate the PMF anisotropic stress, 
which therefore sources both adiabatic scalar and tensor perturbations \cite{Giovannini:2004aw,Lewis:2004ef,Finelli:2008xh,Paoletti:2008ck,Shaw:2009nf}. 
When neutrinos decouple ($ \eta \geq \eta_\nu $), they also produce an anisotropic stress that compensates the magnetic one leading to isocurvature-like perturbations, 
the so-called compensated modes \cite{Giovannini:2004aw,Lewis:2004ef,Finelli:2008xh,Paoletti:2008ck,Shaw:2009nf,Kojima:2009gw}. 

The amplitude of these modes is set by the comoving curvature perturbation $ \zeta $. 
The presence of a PMF sources the growth of $ \zeta $ before neutrino decoupling through the anisotropic stress $ \Pi_B $ \cite{Shaw:2009nf}. 
Once neutrino compensation on the PMF anisotropic stress is effective, the growth of $ \zeta $ ceases. 
For scalar perturbations the final form of $ \zeta $ is \cite{Lewis:2004ef,Shaw:2009nf} 
\begin{equation}
\label{eq:zeta}
\zeta \approx \zeta(\eta_B) - \frac{1}{3}R_\gamma\Pi_B\left[\ln(\eta_\nu/\eta_B) + \(\frac{5}{8R_\nu} - 1 \) \right]\,\,,
\end{equation}
where $ \zeta(\eta_B) $ is the comoving perturbation at the time of PMF generation, 
$ (\Pi_{B})^i_j $ is the magnetic dimensionless anisotropic stress, and $ R_i $ represents the ratio between the total density and the density of the species $ i $. 

From Eq.~\eqref{eq:zeta} we see that there are two main contributions to $ \zeta $. 
The first contribution which has amplitude proportional to the product $ \Pi_B\ln(\eta_\nu/\eta_B)$ is the adiabatic-like passive mode 
(which, unlike the standard adiabatic mode, has non-Gaussian statistics). 
This mode grows logarithmically in time when $\eta_B \leq \eta \leq \eta_\nu$ and then freezes on super-horizon scales after neutrino decoupling. 
The value of $\eta_B$ cannot be defined unless the generation mechanism of the PMF is known. 
In the following, we will allow $\eta_\nu/\eta_B$ to range from $10^6$ to $10^{17}$ \cite{Zucca:2016iur}, 
corresponding to a energy scale of PMF generation between $10^3\,\rm GeV$ and $10^{14}\,\rm GeV$.\footnote{{For inflationary magnetogenesis, 
this corresponds to considering instantaneous reheating at energies $T^4_{\rm reh}\approx H^2_{\rm inf}M^2_{\rm P}/10$ 
between $T_{\rm reh}=10^3\,\rm GeV$ and $T_{\rm reh}=10^{14}\,\rm GeV$ \cite{Bonvin:2013tba}.}} 
From \eq{zeta}, we see that increasing $\eta_\nu/\eta_B$ leads to an increase in the amplitude of the passive mode: 
as we are going to discuss in the following (see Section \ref{sec:4}), 
on large scales this effect will be degenerate with changing the amplitude of the primordial magnetic field. 
The second scalar mode is the so-called compensated mode: 
it is proportional to $ \Pi_B $ but it also has a dependency on the magnetic contributions to the radiation density contrast, $ \Delta_B $, through the ratio $ R_\gamma/R_\nu $. 
This is a isocurvature-like mode sourced by the residual PMF stress-energy after neutrino compensation. 

An expression similar to Eq.~\eqref{eq:zeta} can be derived for tensor perturbations, 
with $ \Pi_B $ replaced by the tensor part of the PMF anisotropic stress. There will be both passive and compensated tensor modes: 
however, compensated tensor modes are small in amplitude and can be safely neglected \cite{Lewis:2004ef}. 
Finally, since vector perturbations rapidly decay when not sourced, there are no passive vector modes. 
Nevertheless, there is a compensated vector mode proportional to the vector part of the anisotropic stress. 

In conclusion, the CMB anisotropy spectra will receive contributions from four modes in total: 
a passive and a compensated scalar mode, a passive tensor mode, and a compensated vector mode. 
In the following sections we briefly discuss what are our fiducial parameters for the PMF power spectrum, 
and see what are the imprints of these four modes on CMB angular spectra.

\subsection{Impact of PMFs on CMB spectra and fiducial model}
\label{sec:impact_fiducial}

\noindent We use the publicly available code MagCAMB\footnote{\url{https://alexzucca90.github.io/MagCAMB/}} \cite{Zucca:2016iur}, 
which is based on a modified version of the Boltzmann integrator CAMB \cite{Lewis:1999bs}, 
to compute the contributions to the CMB angular spectra of the four magnetic modes. The fiducial cosmological model that we are going to use for the forecasts in this paper is 
a flat $\Lambda$CDM model with parameters compatible with the recent Planck 2015 constraints 
\cite{Ade:2015xua,Ade:2015cva,Aghanim:2016yuo}. 
Most importantly, since our aim is to investigate the impact of a PMF on the determination of the tensor-to-scalar ratio $r$ from inflationary GWs, 
we have assumed a fiducial model with \emph{no} primordial GWs (\ie we fix $r=0$), but with a non-zero PMF amplitude. 
More precisely, we choose a PMF amplitude $\Bmpc=1.08\,\mathrm{nG}$ and a time $\eta_\nu/\eta_B = 10^{12}$ of generation of the PMF 
(compatible with the current Planck bounds coming from CMB anisotropies \cite{Ade:2015cva}): 
the reason for these choices is explained in detail at the end of this section. 
Besides, as discussed in the previous section, we choose a spectral index $n_B=-2.9$ for the PMF spectrum. 
For convenience of the reader, we list the values of the cosmological parameters and of the parameters describing the PMF in Tab.~\ref{tab:par}.

\begin{table}
\begin{center}
\begin{tabular}{ccccccc}
$\Omega_{b}h^2$&$\Omega_{c}h^2$&$\tau$&$n_s$&$100\,\theta_{\rm MC}$&$\ln\left[10^{10}A_{s}\right]$&$r$\\
$0.02225$ & $0.01198$ & $0.055$	&$0.9645$ &$1.04077$ &$3.094$&$0$\\
\toprule
\horsp
& &	$\Bmpc \ [\mathrm{nG}]$&$\log_{10}\left(\eta_{\nu}\ \eta_{B}\right)$&$n_{B}$ & &\\
& & $1.08$ &$12$ &$-2.9$ & &
\end{tabular}
\end{center}
\caption{Cosmological (top) and magnetic field (bottom) parameters assumed for the fiducial model. 
The fiducial values for the PMF parameters are within the $95 \%$ C.L. limits 
from current CMB experiments such Planck \cite{Ade:2015cva} and Planck+SPT \cite{Zucca:2016iur}
(see \cite{Paoletti:2010rx,Paoletti:2012bb} for constraints from pre-Planck data).}
\label{tab:par}
\end{table}

\begin{figure*}
\centering
\begin{tabular}{cc}
\includegraphics[width=0.48\columnwidth,keepaspectratio=true]{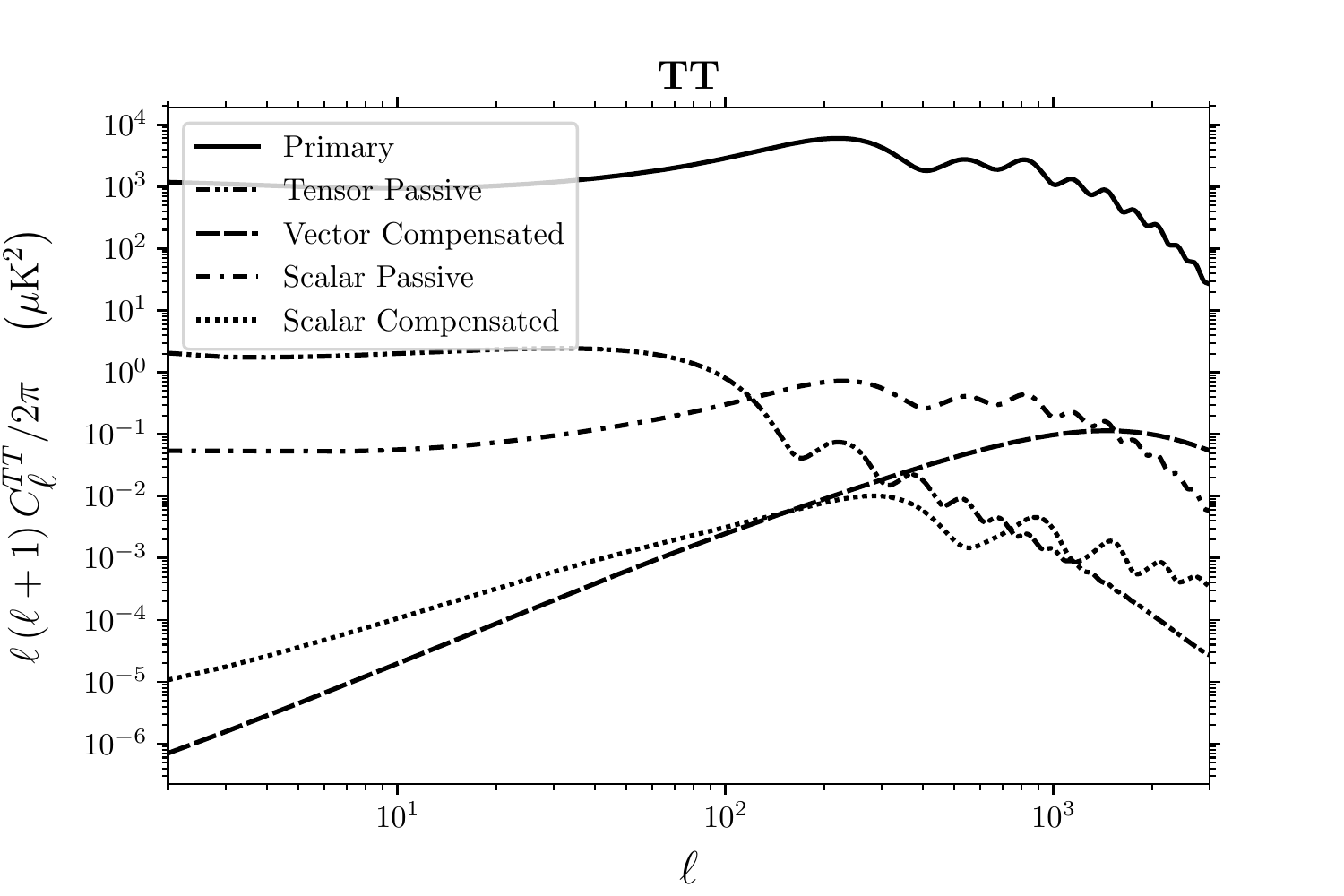} &
\includegraphics[width=0.48\columnwidth,keepaspectratio=true]{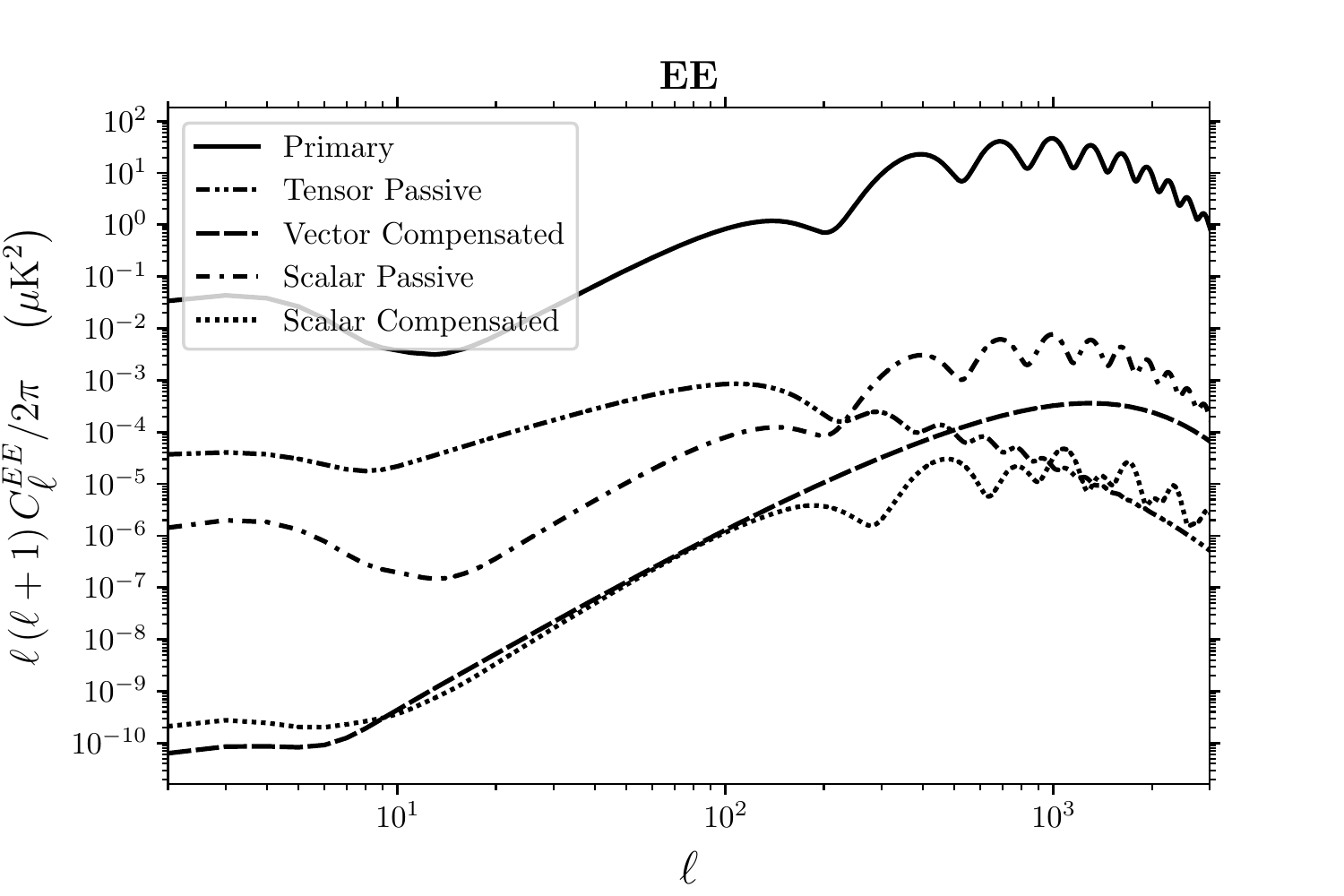} \\
\includegraphics[width=0.48\columnwidth,keepaspectratio=true]{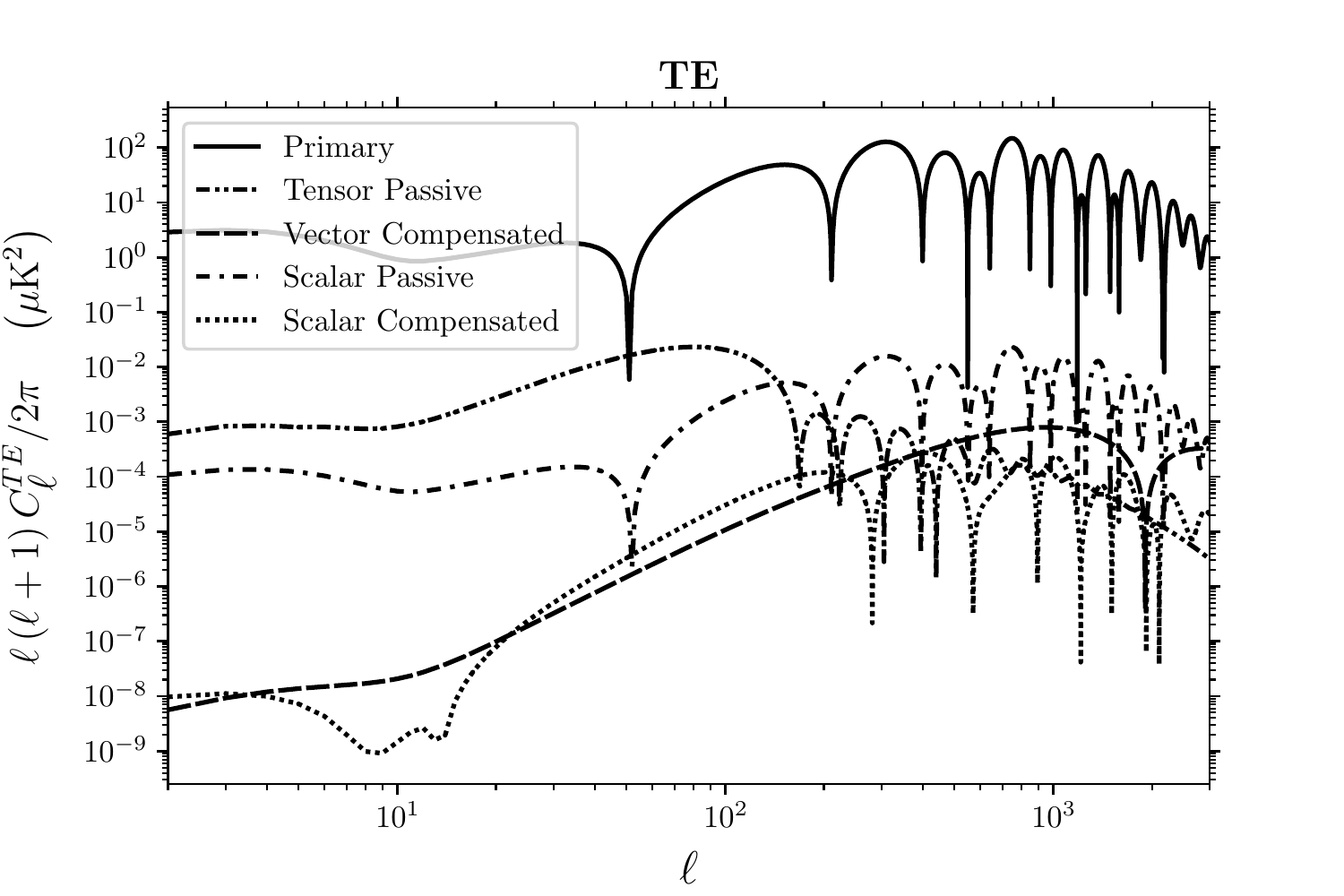} &
\includegraphics[width=0.48\columnwidth,keepaspectratio=true]{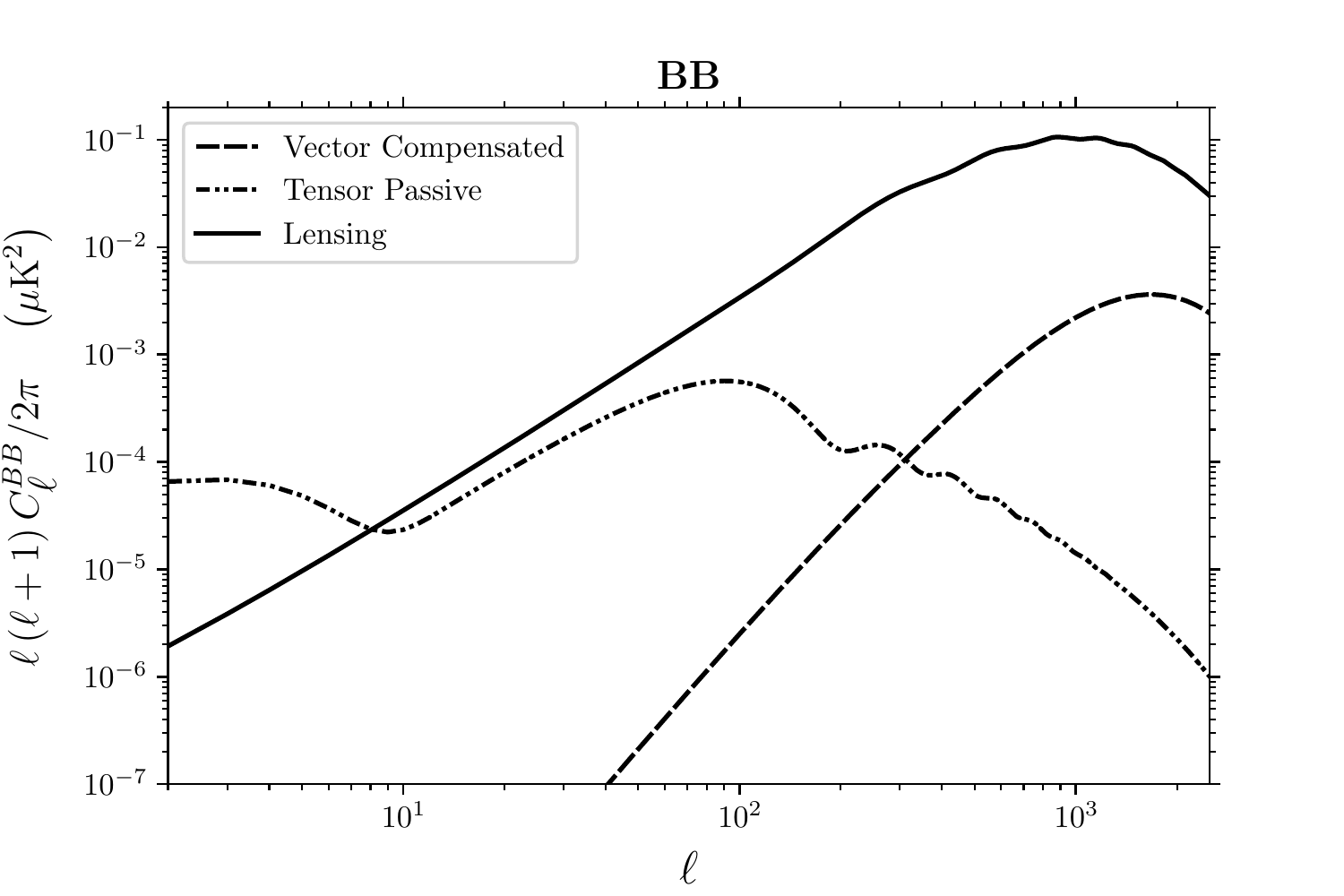} 
\end{tabular}
\caption{Comparison between primary and magnetic contributions to CMB angular correlation functions: 
the spectra have been computed with MagCAMB \cite{Zucca:2016iur}. }
\label{fig:fiducial_vs_primary_Cls}
\end{figure*}

\begin{figure*}
\centering
\includegraphics[width=0.7\columnwidth,keepaspectratio=true]{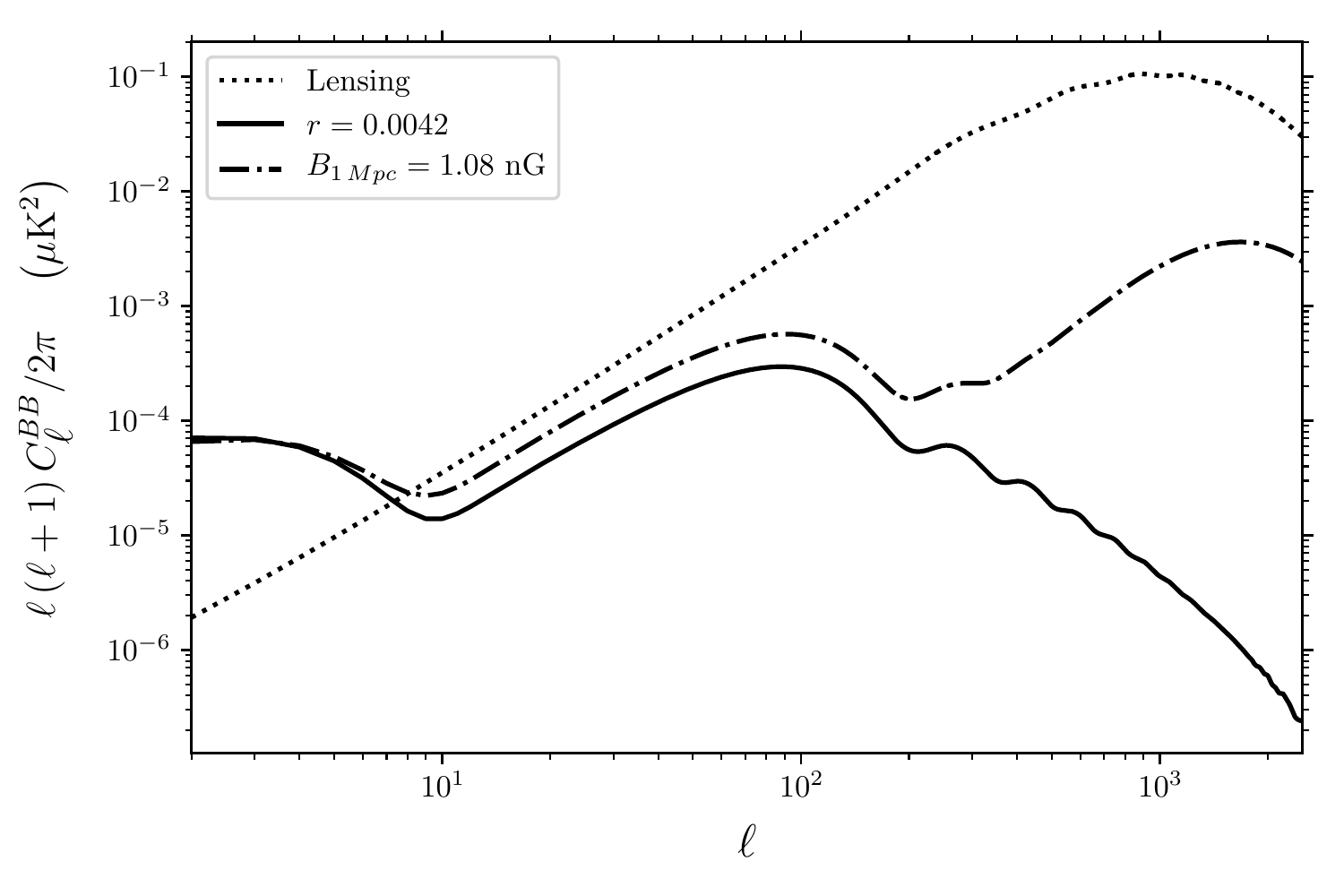}
\caption{Comparison between the (tensor passive + vector compensated) $B$-mode angular power spectrum from PMFs (for $\Bmpc=1.08\,\mathrm{nG}$ and 
$\eta_\nu/\eta_B = 10^{12}$) and the one from inflationary tensor modes (with tensor-to-scalar ratio equal to $0.0042$). 
For this particular choice of parameters, we see that the inflationary and magnetic contributions are degenerate at very large scales, 
where both are about two orders of magnitude larger than lensing $B$-modes.} 
\label{fig:fiducial_plot}
\end{figure*}

Fig. \ref{fig:fiducial_vs_primary_Cls} shows the $TT$, $EE$, $TE$ and $BB$ angular spectra for our fiducial model (Tab.~\ref{tab:par}). 
We see that PMFs have a significant effect mainly on the $BB$ power spectrum. 
On large scales, $\ell\lesssim 100$, the passive tensor mode of the PMF gives a scale-dependence similar to that from inflationary GWs. 
On small scales, magnetic vector perturbations dominate, and lead to an increase in power: 
this feature is not shared by inflationary tensor modes, 
and we expect that it will allow to break the degeneracy between the two mechanisms. 
We also notice that $B$-modes from lensing have a larger amplitude than those from the compensated vector mode: 
however, their scale dependence is different, so we can expect to be able to disentangle it. 

In Fig.~\ref{fig:fiducial_plot}, instead, we plot the sum of the tensor passive and vector compensated contributions to the $B$-mode 
angular spectrum, together with lensing $B$-modes and the prediction for $C^{BB}_\ell$ given 
a tensor-to-scalar ratio $r=0.0042$. This is the prediction of the Starobinsky $R^2$ model \cite{Starobinsky:1980te} for $N_\star\approx53$ 
($N_\star$ being the number of $e$-folds of the observable part of the inflationary epoch), 
and is the main target of upcoming CMB experiments \cite{Abazajian:2016yjj,Finelli:2016cyd}. 
We see that our fiducial values $\Bmpc=1.08\,\mathrm{nG}$ for the magnetic field amplitude and 
$\eta_\nu/\eta_B = 10^{12}$ for the time of PMF generation 
give a large-scale $B$-mode spectrum very similar to that of the Starobinsky model. 
We take, then, these values of $\Bmpc$ and $\eta_\nu/\eta_B$ as a case-study, 
using them to show how an unresolved PMF component can bias the constraints on the theoretically motivated class of 
inflationary models known as $\alpha$-attractors \cite{Kallosh:2013yoa}. 

Before proceeding, we comment on the degeneracy between $\Bmpc$ and $\eta_\nu/\eta_B$: 
since the amplitude of the tensor passive mode, as that of the scalar mode, is proportional to $\eta_\nu/\eta_B$ \cite{Shaw:2009nf}, 
we expect that increasing this parameter will result in more power in the $BB$ spectrum at low $\ell$. 
More precisely, for nearly scale-invariant PMF spectra the contribution of the tensor passive mode to $C^{BB}_\ell$ scales as 
$C^{BB,\rm passive}_\ell\sim\Bmpc^4\big[\ln(\eta_\nu/\eta_B)\big]^{2}$. 
Therefore, we could have equivalently reproduced the large-scale behavior of the Starobinsky model 
by choosing a larger $\Bmpc$ and a smaller $\eta_\nu/\eta_B$. 
We stress, however, that for us the choice of the fiducial values of these parameters is not important: 
what is relevant is how a contribution to the $B$-mode power spectrum from PMFs could 
lead to a false claim of a detection at the level $r\approx\num{d-3}$. 
Moreover, choosing a larger $\Bmpc$ and a smaller $\eta_\nu/\eta_B$ 
would lead to a smaller contribution from the compensated vector mode. 
Therefore, breaking the degeneracy would be even more difficult for experiments that have access only to large scales. 
This makes our choice of parameters the most conservative one.\footnote{Of course, this point can be turned around. 
We can have the same large-scale tensor power by taking a larger $\eta_\nu/\eta_B$ 
and a smaller $\Bmpc$: in that case the contribution of the compensated vector mode 
could be large enough to be observable also on large scales. 
However, we stress that our point is that there is always a region in the currently available parameter space 
where the degeneracy cannot be broken unless we have access to small scales. }

\section{Method}
\label{sec:3}

\noindent In this section we briefly illustrate the method we adopted to derive our forecasts for future CMB experiments. 
We follow the same procedure (now standard practice) used in \cite{Perotto:2006rj}. We produce synthetic realizations of future data given by
\begin{equation}
\label{fiducial}
\hat{C}_\ell = C_\ell|_\text{fid} + N_\ell\,\,.
\end{equation}
On the right-hand side, the $C_\ell|_\text{fid}$ are the angular power spectra of the fiducial model in $\mu\mathrm{K}^2$ and 
$N_\ell = w^{-1}\exp(\ell(\ell+1)\theta^2/8\ln2)$ 
gives the experimental noise, where $w^{-1/2}$ is the experimental power noise expressed in 
$\mu \mathrm{K} \cdot \text{rad}$ and $\theta$ is the experimental FWHM angular resolution in radians 
(we assume that pixel noise is uniform and uncorrelated). 
We have considered several future experiments with technical specifications listed in Tab.~\ref{tab:spec}. 
More specifically we consider the PIXIE \cite{Kogut:2011xw}, LiteBIRD \cite{Suzuki:2018cuy} and CORE-M5 \cite{Delabrouille:2017rct} satellite missions, 
and the Stage-3 (see e.g. \cite{Allison:2015qca}) and CMB-S4 \cite{Abazajian:2016yjj} ground-based experiments.\footnote{Most of these experiments are still in 
the stage of a proposal. The above list should therefore be considered as an illustration of what a future CMB experiments could achieve.} 
The simulated data are then compared with theoretical $C_{\ell}$ obtained from MagCAMB. 
The likelihood function employed in our forecasts is the inverse Wishart \cite{Perotto:2006rj}, commonly known as ``exact likelihood''.

\begin{table}
\begin{center}
\begin{tabular}{lccccc}
\toprule
\horsp
Experiment 
&Beam 
&Power noise 
&$\ell_{\rm max}$& $\ell_{\rm min}$& $f_{\rm sky}$\\
& 
[\footnotesize{$\mathrm{arcmin}$}] 
&
[\footnotesize{$\mu\mathrm{K}\cdot\mathrm{arcmin}$}]
& & &\\[0.2em]
\hline
\hline
\morehorsp
PIXIE & $96$ 
& $3.0$ & $500$&$2$&$0.7$\\ 
LiteBIRD & $30$
& $3.2$& $3000$&$2$&$0.7$\\ 
CORE-M5 & $3.7$
& $2.0$& $3000$&$2$&$0.7$\\ 
Stage-3 (Deep) & $1$
& $4$ & $3000$&$50$&$0.06$\\
Stage-3 (Wide) & $1.4$
& $8$ & $3000$&$50$&$0.4$\\
CMB-S4 & $3$
& $1$& $3000$&$5$&$0.4$\\
\bottomrule
\end{tabular}
\end{center}
\caption{Experimental specifications for the several configurations considered in the forecasts. 
The power noise is defined as $w^{-1/2}=\sqrt{4\pi\sigma^2/N}$, where $\sigma$ is the r.m.s. noise in each of the $N$ pixels. 
We quote the power noise for temperature, and assume that for polarization it is simply enhanced by a factor of $\sqrt{2}$.}
\label{tab:spec}
\end{table}

In the following, we sample the likelihood using a MagCAMB-compatible \cite{Zucca:2016iur} version of the Monte Carlo Markov Chain code 
{CosmoMC}\footnote{\url{http://cosmologist.info}}~\cite{Lewis:2002ah}, based on the Metropolis-Hastings algorithm with 
chains convergence tested by the Gelman and Rubin method. 

Since we assumed no correlation between primary adiabatic and magnetic modes, 
both theory and fiducial $C_\ell$ are obtained simply adding together magnetic and non-magnetic contributions, i.e. 
\begin{equation}
\label{sum_spectra}
C_\ell = C_\ell^{\rm primary} + C_\ell^{\rm passive} + C_\ell^{\rm compensated}\,\,.
\end{equation}

We also study the impact of delensing on future constraints, to understand if this procedure can help in breaking the degeneracy 
discussed in Section \ref{sec:impact_fiducial}. 
We subtract from the total signal the lensed CMB $B$-modes using the ``CMB$\times$CMB'' delensing procedure already proposed in 
\cite{Smith:2010gu,Errard:2015cxa}. 
For each future experiment considered in this paper, 
we rescale the $BB$ power spectrum by using the corresponding delensing factor $\alpha$ \cite{Smith:2010gu,Sherwin:2015baa,Errard:2015cxa}, 
defined as the ratio between the total delensed $BB$ power spectrum 
and the total original lensed one (see \cite{Errard:2015cxa} for a detailed description of the delensing procedure). 
The values of the delensing factor $\alpha$ for the different experiments of Tab.~\ref{tab:spec} are collected in Tab.~\ref{tab:alpha}.
When delensing is included, we do not carry out a full exploration of the parameter space. 
We sample the exact likelihood only for $B$-modes, fixing all parameters apart from $r$ and $\Bmpc$.

\begin{table}
\begin{center}
\begin{tabular}{cccccc}
PIXIE & LiteBIRD & CORE-M5 & Stage-3 (Deep) & Stage-3 (Wide) & CMB-S4\\
\toprule
\horsp
$1$	&$0.94$	&$0.37$	&$0.56$	&$0.79$	&$0.25$
\end{tabular}
\end{center}
\caption{Delensing factor $\alpha$ for the various experiments described in Tab.~\ref{tab:spec}.} 
\label{tab:alpha}
\end{table}

\section{Results}
\label{sec:4}

\noindent In what follows we analyze the simulated datasets for the fiducial of Tab.~\ref{tab:par} and the experiments of Tab.~\ref{tab:spec}. 
For $r$ and $\Bmpc$ we use a linear prior in the range $[0,3]$ and $[0,5]$, respectively. 
We instead sample logarithmically $\eta_{\nu}/\eta_B$ in the range $[10^6,10^{17}]$. 
The inflationary tensor spectral index is given by the consistency relation $n_t=-r/8$ 
while we consider only nearly scale-invariant PMFs with $n_B=-2.9$.

\subsection{Results from MCMC}
\label{sec:MCMC}

\noindent Let us start from the MCMC analysis. As a first step, we analyze the ability of future/planned CMB experiments (listed in Tab.~\ref{tab:spec}) 
of recovering our fiducial values for the PMF parameters. 

The constraints on $\Bmpc$ from our selection of future experiments are reported in Tabs.~\ref{future-1}, \ref{future-2}. 
In the first row of these tables we report the results from our analysis without variation in $\eta_\nu/\eta_B$. 

In this case, we see that large satellite experiments as PIXIE and LiteBIRD can recover $\Bmpc$ with a very good precision 
of about $\sigma(\Bmpc)\sim 0.06\,\text{-}\, 0.03\,\mathrm{nG}$. 
Conversely, a satellite mission as CORE-M5, with improved angular solution respect to PIXIE or LiteBIRD, 
would measure the PMF amplitude with an excellent accuracy of $\sigma(\Bmpc)\sim 0.02\,\mathrm{nG}$. 
The same accuracy can be achieved by the CMB-S4 experiment. 
However it is important to note that we assumed a quite optimistic value of $\ell_{\rm min}=5$ for CMB-S4. 
It may be possible that the final $\ell_{\rm min}$ will shift towards higher values given a more limited scanning strategy due to, 
for example, shorter observation time and high frequency foregrounds. 
The Stage-3 experiment in both configurations ``wide'' or ``deep'' will provide much weaker constraints. 
This is due to the fact that the Stage-3 experiment is less sensitive to large-scale $B$-modes. 

It is interesting to note that while experiments as CORE-M5 and CMB-S4 can constrain a PMF with amplitude of $\Bmpc=1.08\,\mathrm{nG}$ 
with a $\sigma(\Bmpc)\sim 0.02\,\mathrm{nG}$ precision, the upper limit on $\Bmpc$ achievable from CMB-S4 in case of no PMF is 
$\Bmpc <0.52\,\mathrm{nG}$ at $68 \%$ C.L. (see e.g. \cite{Pogosian:2018vfr}). 
This is due to the fact that, at fixed $\eta_\nu/\eta_B$, the amplitude of the $B$-mode polarization from PMF scales as $C_{\ell} \sim \Bmpc^4$. 
This means that the CMB will be able to strongly constrain a PMF, 
if detected, but also that upper limits on $\Bmpc$ will be only marginally improved by future experiments 
(by a factor four in case of CMB-S4 if we compare with the upper limit of $2\,\mathrm{nG}$ from Planck). 

These constraints are obtained under the assumption of a perfect knowledge of the time of PMF generation, 
and this is obviously not a realistic case. 
As we see from the second row of both Tabs.~\ref{future-1}, \ref{future-2}, letting also $\eta_\nu/\eta_B$ free to vary 
weakens the constraints on $\Bmpc$. The constraints on $\Bmpc$ are relaxed by a factor $\sim 4\,\text{-}\, 5$ for the PIXIE and LiteBIRD experiments, 
a factor $\sim 2$ for CORE-M5 and by $\sim 50 \%$ for CMB-S4. 
Moreover, the posterior of $\eta_\nu/\eta_B$ is tighter than the prior essentially just with the CORE-M5 and CMB-S4 experiments. 

Let us now see how the degeneracy between PMFs and a possible primordial GW component can be resolved. 
To this goal, we perform three different additional analyses: 
\begin{itemize}
\item in the first analysis, in contrast to our fiducial model, we wrongly assume no PMF and we let only $r$ free to vary 
with a tensor spectral index $n_t$ given by the inflationary consistency relation $n_t=-r/8$. 
Here we quantify how the assumption of no PMF could bias the determination of $r$ and provide a misleading first detection of primordial GWs; 
\item in the second analysis we consider as free parameters both $r$ and a $\Bmpc$, but again we fix the ratio $\eta_\nu/\eta_B$, 
to understand if a CMB experiment could discriminate between inflationary GWs 
and a PMF when the latter is described just by one parameter; 
\item in the third analysis we vary $r$, $\Bmpc$ and $\eta_\nu/\eta_B$. 
As noticed in \cite{Pogosian:2018vfr} and discussed in Section \ref{sec:impact_fiducial}, 
since the time ratio determines the amplitude of the passive tensor modes, it is mostly degenerate with $r$ in the $B$-mode polarization.
\end{itemize}

\begin{table*}
\begin{center}\footnotesize
\scalebox{1}{\begin{tabular}{lccccccc}
\toprule
\horsp
&PIXIE	&LiteBIRD		&CORE-M5
\\ 		 
\hline
\morehorsp
$\log_{10}(\eta_{\nu}/\eta_B)=12$	&	&	&	
\\ \morehorsp

$\Bmpc$ ($\mathrm{nG}$)		&$1.07\pm0.10$&$1.078^{+0.034}_{-0.028}$	&$1.080^{+0.037}_{-0.038}$	
\\

\hline
\hline
\morehorsp

$\log_{10}(\eta_{\nu}/\eta_B)$ free	&	&	&
\\ \morehorsp

$\Bmpc$ ($\rm nG$)	&$1.16^{+0.35}_{-0.31}$	&$1.15^{+0.34}_{-0.28}$	&$1.07^{+0.10}_{-0.11}$
\\
$\log_{10}(\eta_{\nu}/\eta_B)$	&unconstrained		&unconstrained		&$12.4^{+2.9}_{-2.5}$
\\

\bottomrule
\end{tabular}}
\caption{$95 \%$ C.L. constraints on $\Bmpc$ and $\log_{10}(\eta_{\nu}/\eta_B)$ for the PIXIE, LiteBIRD and CORE-M5 experiments. 
These runs are assume that inflationary gravitational waves are absent, \ie $r=0$.} 
\label{future-1}
\end{center}
\end{table*}

\begin{table*}
\begin{center}\footnotesize
\scalebox{1}{\begin{tabular}{lccccccc}
\toprule
\horsp
	
&Stage-3 (Deep)	&Stage-3 (Wide)	&CMB-S4\\ 		 
\hline
\morehorsp
$\log_{10}(\eta_{\nu}/\eta_B)=12$	
&	&	&	\\ \morehorsp

$\Bmpc$ ($\mathrm{nG}$)		
&$<1.2$&$<1.2$	&$1.079 \pm 0.020$\\

\hline
\hline
\morehorsp

$\log_{10}(\eta_{\nu}/\eta_B)$ free & &	&	\\ \morehorsp

$\Bmpc$ ($\rm nG$) &$<1.3$	&$<1.3$	&$1.074^{+0.061}_{-0.065}$ \\
$\log_{10}(\eta_{\nu}/\eta_B)$	&unconstrained		&unconstrained	&$12.2\pm1.8$ \\

\bottomrule
\end{tabular}}
\caption{As in Tab.~\ref{future-1} for the Stage-3 (deep and wide configurations) and CMB-S4 experiments. } 
\label{future-2}
\end{center}
\end{table*}

\begin{figure*}
\centering
\begin{tabular}{cc}
\includegraphics[width=0.48\columnwidth,keepaspectratio=true]{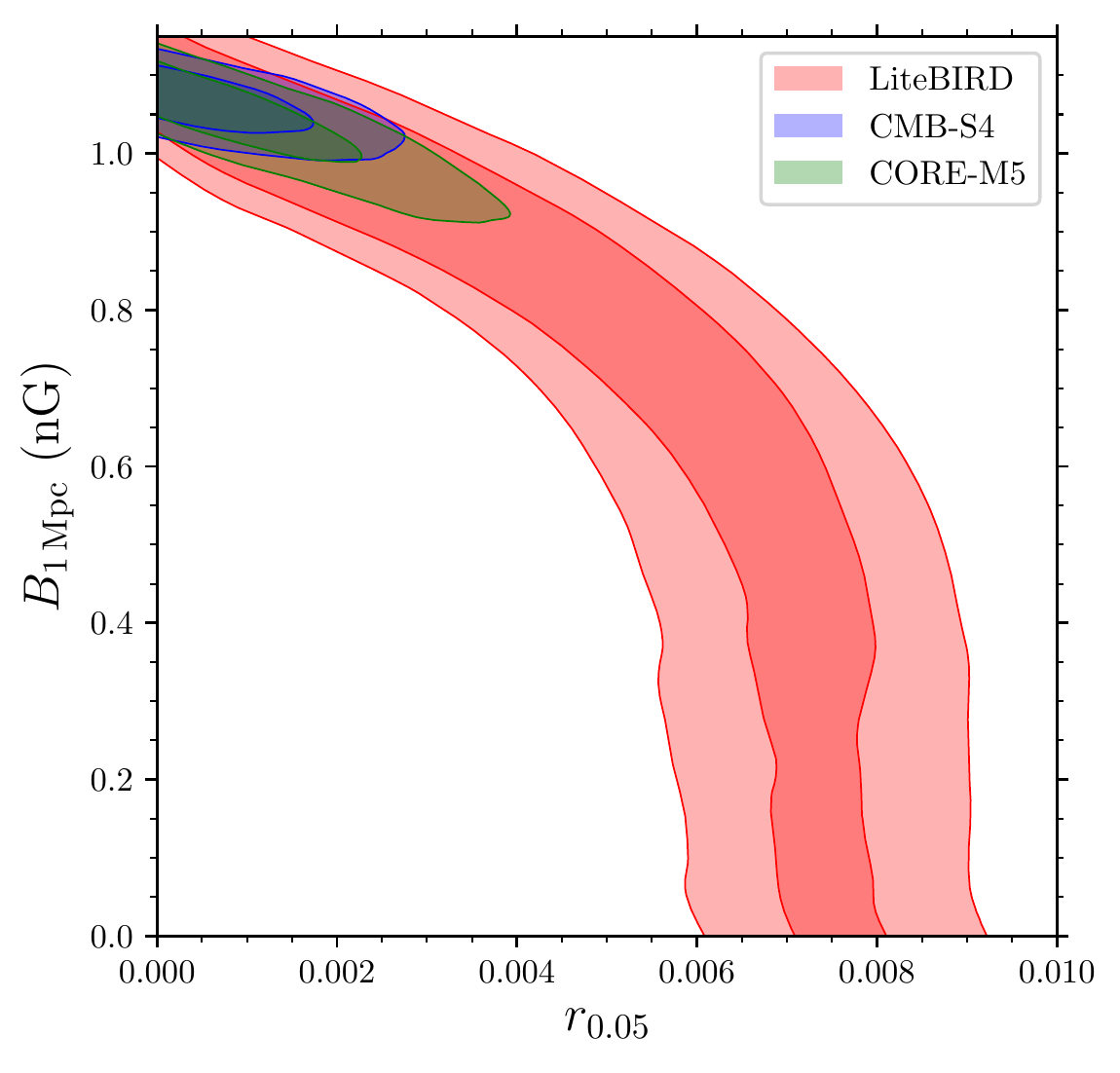} &
\includegraphics[width=0.48\columnwidth,keepaspectratio=true]{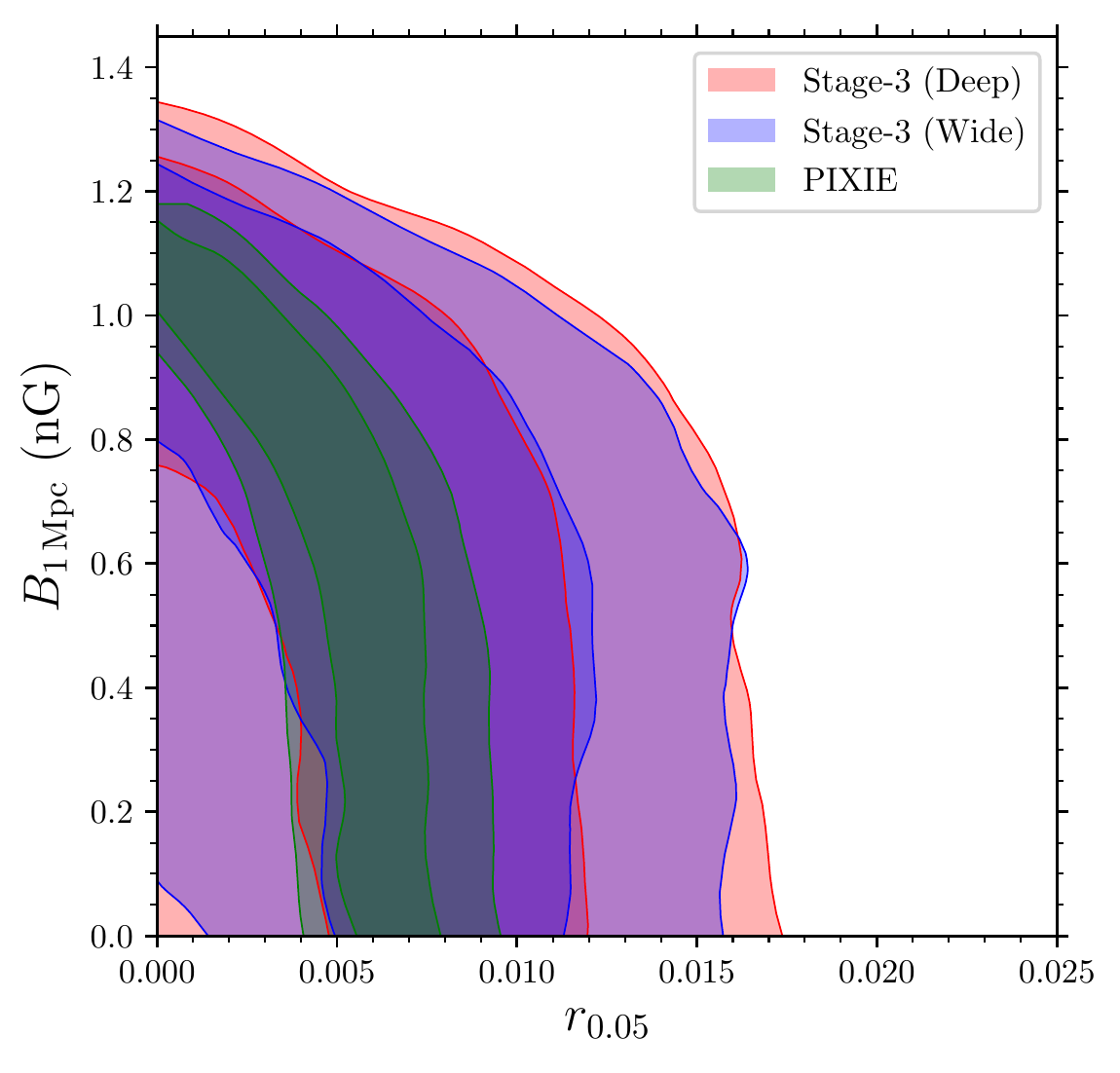}
\end{tabular}
\caption{Forecasted constraints in the $\Bmpc$ vs. $r$ plane from MCMC 
for LiteBIRD and CMB-S4 (left panel) and for PIXIE, Stage-3, and CORE-M5 (right panel). 
The fiducial model has $\Bmpc=1.08\,\mathrm{nG}$, $\eta_\nu/\eta_B=10^{12}$ and $r=0$. 
Clearly the LiteBIRD, PIXIE and Stage-3 experiments are unable to distinguish the PMF from inflationary GWs. 
On the contrary, CMB-S4 and CORE-M5 can break the degeneracy thanks to better sensitivity to small scale $B$-modes.}
\label{fig:litebirdvss4}
\end{figure*}

\begin{table*}
\begin{center}\footnotesize
\scalebox{1}{\begin{tabular}{lccccccc}
\toprule
\horsp
&PIXIE	&LiteBIRD		&CORE-M5
\\		 

\hline
\morehorsp

$r$	&$0.0065^{+0.0029}_{-0.0028}$	&$0.0073^{+0.0018}_{-0.0017}$	&$0.0072 \pm 0.0011$	\\

\hline
\hline
\morehorsp

$r$	&$0.0050^{+0.0037}_{-0.0047}$	&$<0.0082$	&$ < 0.0031$\\ \morehorsp

$\Bmpc$ ($\mathrm{nG}$)		&$<1.1$	&$<1.1$	&$1.034^{+0.079}_{-0.096}$\\

\hline
\hline
\morehorsp

$r$	&$ <0.0083$	&$0.0051^{+0.0035}_{-0.0046}$		&$<0.0057$\\ \morehorsp

$\Bmpc$ ($\mathrm{nG}$)		&$< 1.2$	&$<1.3$	&$ 1.06^{+0.11}_{-0.12}$ \\ \morehorsp

$\log_{10}(\eta_{\nu}/\eta_B)$	&unconstrained		&unconstrained		&$<13$\\

\hline
\hline
\end{tabular}}
\caption{$95 \%$ C.L. constraints on $\Bmpc$, $\log_{10}(\eta_{\nu}/\eta_B)$ and $r$ for the PIXIE, LiteBIRD and CORE-M5 experiments. 
Notice that these are one-dimensional marginalized constraints: 
given the strong degeneracy between $r$ and $\Bmpc$, 
the detection of $r$ when both parameters are varied is not significant.} 
\label{tabledeg-1}
\end{center}
\end{table*}

\begin{table*}
\begin{center}\footnotesize
\scalebox{1}{\begin{tabular}{lccccccc}
\toprule
\horsp
	
&Stage-3 (Deep)	&Stage-3 (Wide)	&CMB-S4\\		 

\hline
\morehorsp

$r$	&$ 0.0084^{+0.0079}_{-0.0084}$ &$0.0084^{+0.0075}_{-0.0081}$	&$0.0072 \pm 0.0014$\\

\hline
\hline
\morehorsp

$r$&$<0.015$	&<0.015&$< 0.0022$ \\ \morehorsp

$\Bmpc$ ($\mathrm{nG}$)	&$<1.1$	&$<1.1$	&$1.058^{+0.053}_{-0.055}$ \\

\hline
\hline
\morehorsp

$r$&$<0.015$	&$<0.014$	&$<0.0059$ \\ \morehorsp

$\Bmpc$ ($\mathrm{nG}$)	&$< 1.2$	&$<1.2$	&$1.073^{+0.065}_{-0.069}$ \\ \morehorsp

$\log_{10}(\eta_{\nu}/\eta_B)$&unconstrained		& unconstrained &	$<12$	\\

\hline
\hline
\end{tabular}}
\caption{As in Tab.~\ref{tabledeg-1} for the Stage-3 (deep and wide configurations) and CMB-S4 experiments.} 
\label{tabledeg-2}
\end{center}
\end{table*}

The results of these three analysis are reported in Tabs.~\ref{tabledeg-1}, \ref{tabledeg-2}.
As we can see from the first row of both Tabs.~\ref{tabledeg-1}, \ref{tabledeg-2}, 
analyzing a CMB dataset with the wrong assumption of no PMF and $\Bmpc=0$ could lead to a bias on the tensor-to-scalar ratio $r$. 
An experiment as PIXIE could provide an indication at above three standard deviations for a primordial tensor amplitude of $r\sim 0.0065$. 
LiteBIRD, CORE-M5 and CMB-S4 will provide an even higher evidence for $r>0$, 
with statistical significances that could reach (and also go over) about 10 standard deviations. 
The experimental evidence for $r>0$ that these experiments obtain is therefore completely misleading and based on the wrong assumption of $\Bmpc=0$. 

The next step consists in letting also $\Bmpc$ free to vary and see if these future experiments will be able to discriminate between $r$ and $\Bmpc$. 
The results of this analysis are on the second and third rows of both Tabs.~\ref{tabledeg-1}, \ref{tabledeg-2}. 
As we can see, when $\Bmpc$ is included, the detection for $r>0$ simply disappears or is rather weaker for all the experiments considered. 
Furthermore, for the PIXIE, LiteBIRD and Stage-3 experiments there is also no clear detection for $\Bmpc$. 
What is happening is clear by looking at Figure~\ref{fig:litebirdvss4}: 
a degeneracy is present on the $\Bmpc$ vs. $r$ plane and the experiments are simply unable to discriminate between a 
genuine primordial tensor component from GWs and PMFs. 
For PIXIE and LiteBIRD this is essentially due to the poor experimental angular resolution that does not allow to access small scales, $\ell\gtrsim 1000$, 
where the vector compensated PMF $B$-mode could be detected. 
Indeed, when we consider experiments with better angular resolution as CORE-M5 and CMB-S4 the degeneracy is broken, 
the PMF is well measured and just an upper limit is obtained for $r$. 
For the Stage-3 experiment a degeneracy between $r$ and $\Bmpc$ is also present: 
this is essentially due to the lower experimental sensitivity that does not allow clear detection of the $B$-mode signal of the fiducial model. 
Finally, from the last three rows of Tabs.~\ref{tabledeg-1}, \ref{tabledeg-2} we see that the same conclusions hold when we let also $\eta_\nu/\eta_B$ free to vary.

\subsection{Delensing}

\begin{figure*}
\centering
\includegraphics[width=0.6\columnwidth,keepaspectratio=true]{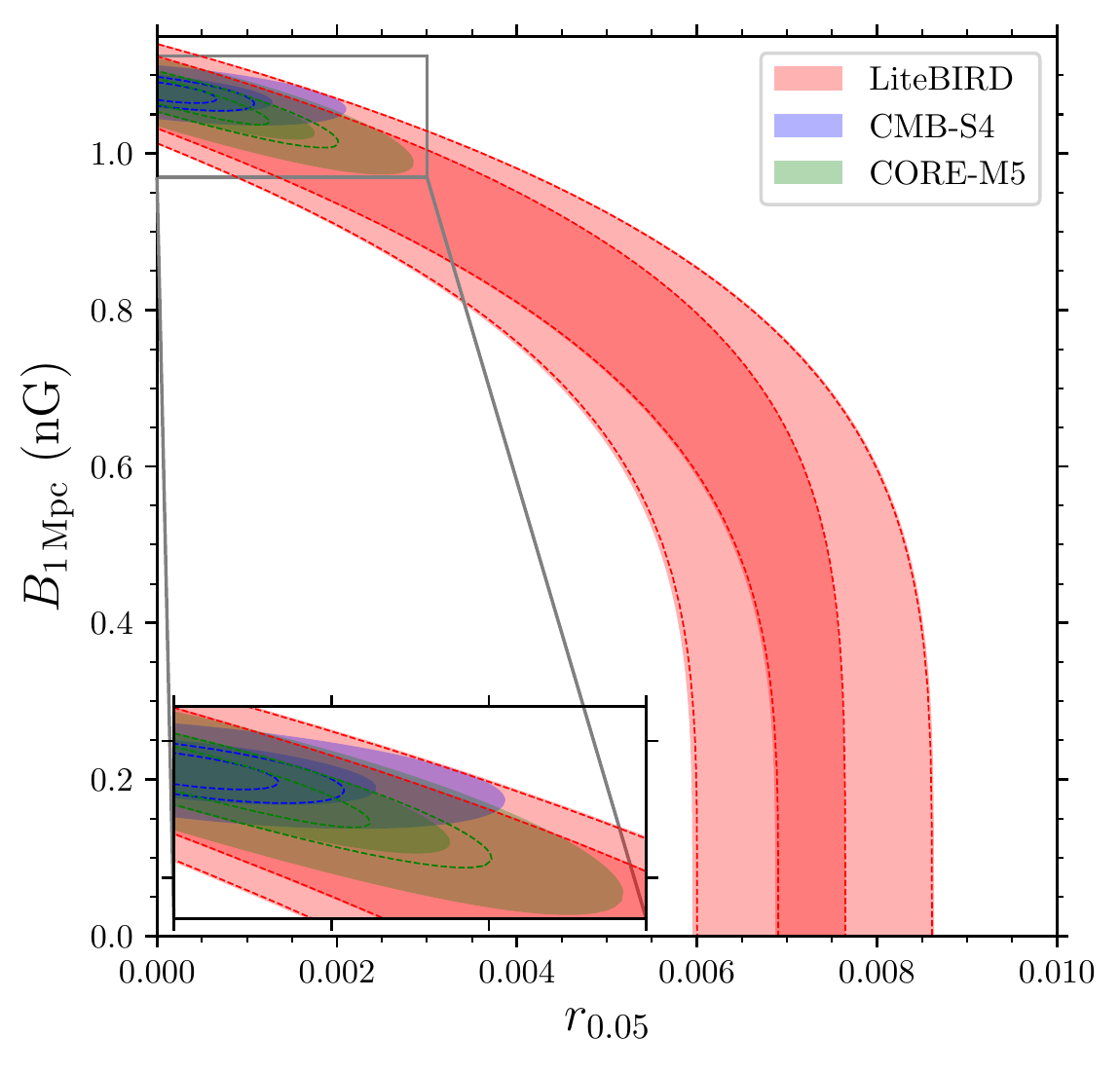}
\caption{Impact of delensing for LiteBIRD, CMB-S4 and CORE-M5: 
dashed lines represent the $68 \%$ C.L. and $95 \%$ C.L. contours after delensing. 
Clearly, the delensing procedure affects the constraints on $r$ and $\Bmpc$ only if the degeneracy between the two is broken.}
\label{fig:r_forecasts}
\end{figure*}

\noindent In this section we briefly study the impact of delensing on the forecasted constraints in the $\Bmpc$ vs. $r$ 
plane from the simulated data for LiteBIRD, CMB-S4 and CORE-M5, using the $BB$ exact likelihood and fixing all parameters apart from $r$ and $\Bmpc$. 
To obtain the theoretical angular spectra we rescale two templates computed for $r=0.1, \Bmpc=0\,\mathrm{nG}$ and $r=0,\Bmpc=1.08\,\mathrm{nG}$ 
(fixing in both cases $\eta_\nu/\eta_B = 10^{12}$). 
We leave $n_t$ fixed to $-0.1/8$, even if $r$ is varied: 
at such low values of $r$ as those probed by LiteBIRD, CMB-S4 and CORE-M5 the error is negligible 
(this is confirmed a posteriori by comparing Fig.~\ref{fig:r_forecasts} with Fig.~\ref{fig:litebirdvss4}).

The $68 \%$ C.L. and $95 \%$ C.L. contours are reported in Fig.~\ref{fig:r_forecasts}. 
First, we notice that even fixing all parameters apart from $\Bmpc$ and $r$ leads to only marginally 
more stringent constraints than those depicted in Fig.~\ref{fig:litebirdvss4} 
(besides, recall that in this case we are not including the information coming from the $TT$, $EE$ and $TE$ spectra). 
Most importantly, we also see that delensing will not help to break the degeneracy between the two parameters for LiteBIRD. 
For CORE-M5 and CMB-S4, instead, delensing would shrink the $95 \%$ C.L. contours of roughly a factor of $2$. 
We did not show the forecasts for Stage-3 or PIXIE in Fig.~\ref{fig:r_forecasts}: 
as for LiteBIRD, also in this case delensing would not help in breaking the degeneracy between $r$ and $\Bmpc$.

\subsection{Importance of small-scale \texorpdfstring{$B$}{B}-mode measurements}

\noindent Finally, we investigate in more detail the differences between an experiment that cannot break 
the degeneracy between $r$ and $\Bmpc$ (\ie LiteBIRD) and one that can (\ie CMB-S4). 

We consider two different models with roughly the same $\chi^2_\mathrm{min}$ for the LiteBIRD simulated dataset, 
but with very different values of $r$ and $\Bmpc$: the model ``$(1)$'' has $(r=1.76\times 10^{-5},\Bmpc=1.064\,\mathrm{nG})$, 
while the model ``$(2)$'' has $(r=7.22\times10^{-3},\Bmpc=7.57\times10^{-2}\,\mathrm{nG})$. 
Then, we expect that for LiteBIRD it will be impossible to distinguish between these two models, 
while CMB-S4 will be able to break the degeneracy between them. 

\begin{figure*}
\centering
\includegraphics[width=0.7\columnwidth,keepaspectratio=true]{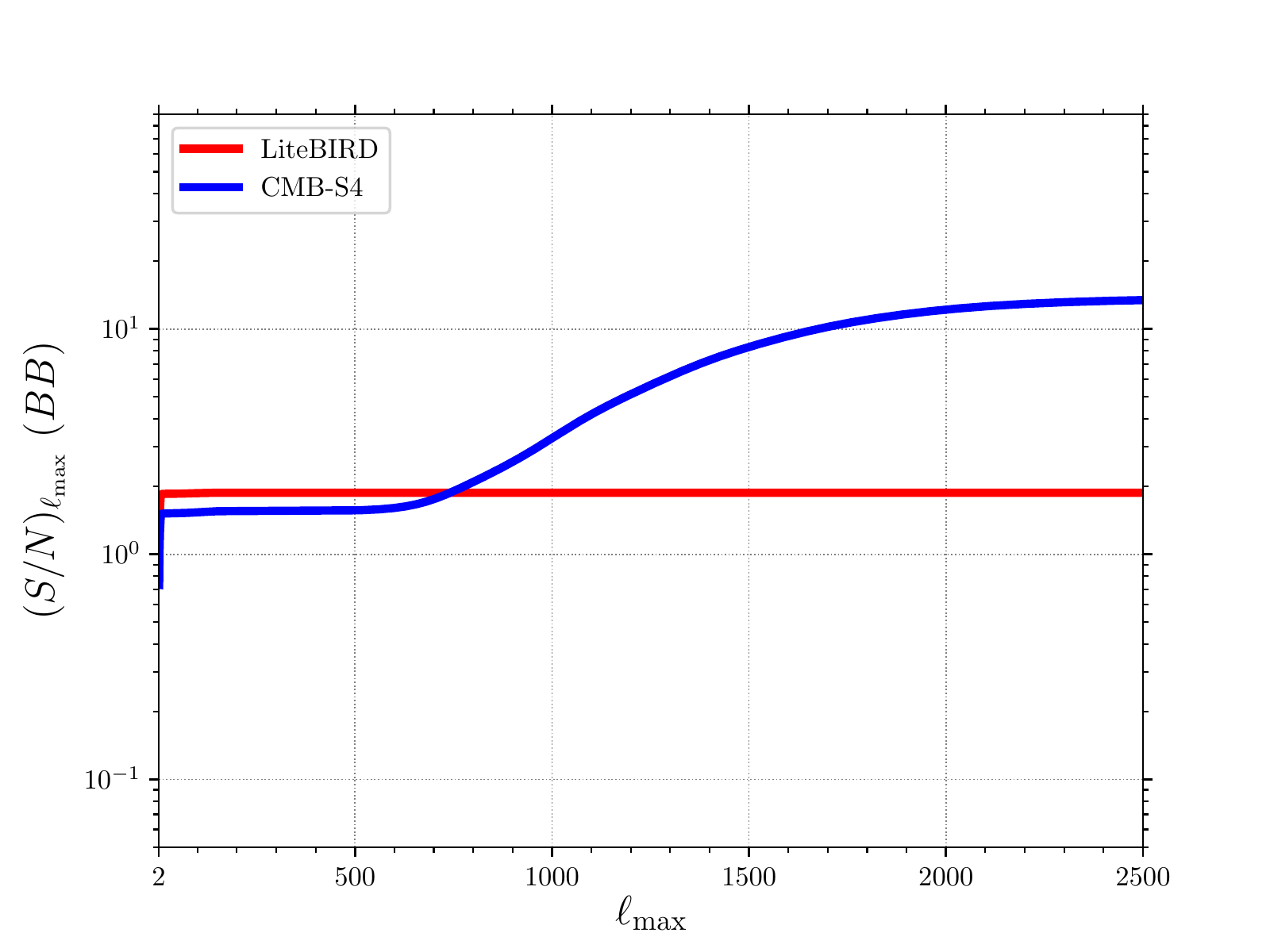}
\caption{$S/N$ of \eq{S_N} at varying $\ell_{\rm max}$ for the LiteBIRD and CMB-S4 experiments.}
\label{fig:SNfig}
\end{figure*}

We can estimate how well the two experiments are able to distinguish model $(2)$ from model $(1)$ by constructing 
a simple $\chi^2_\ell\equiv (\Delta C_\ell/\sigma_\ell)^2$ for the difference $\Delta C_\ell\equiv C_\ell^{(2)}-C_\ell^{(1)}$. 
Thus, assuming uncorrelated multipoles, we can write the cumulative signal-to-noise ratio as 
\begin{equation}
\label{eq:S_N}
\(S/N\)^2_{\ell_{\rm max}}=\sum_{\ell=2}^{\ell_{\rm max}}\frac{\(C_\ell^{(2)}-C_\ell^{(1)}\)^2}{\sigma_\ell^2}\,\,, 
\end{equation}
where we focus on $B$-modes only and $\sigma_\ell$ is given by noise plus cosmic variance, i.e. \cite{Tegmark:1997vs} 
\begin{equation}
\sigma_\ell = \sqrt{\frac{2}{(2\ell+1)f_{\rm sky}}}\,\(C_\ell^{(1)} + N_\ell\)\,\,.
\end{equation}
We plot $S/N$ for varying $\ell_{\rm max}$ in Fig.~\ref{fig:SNfig}: 
we see that for LiteBIRD it remains of order $1$ up to high $\ell_{\rm max}$, 
while for CMB-S4 it becomes of order $10$ at $\ell_{\rm max}\gtrsim 1000$.

\begin{figure*}
\centering
\includegraphics[width=0.6\columnwidth,keepaspectratio=true]{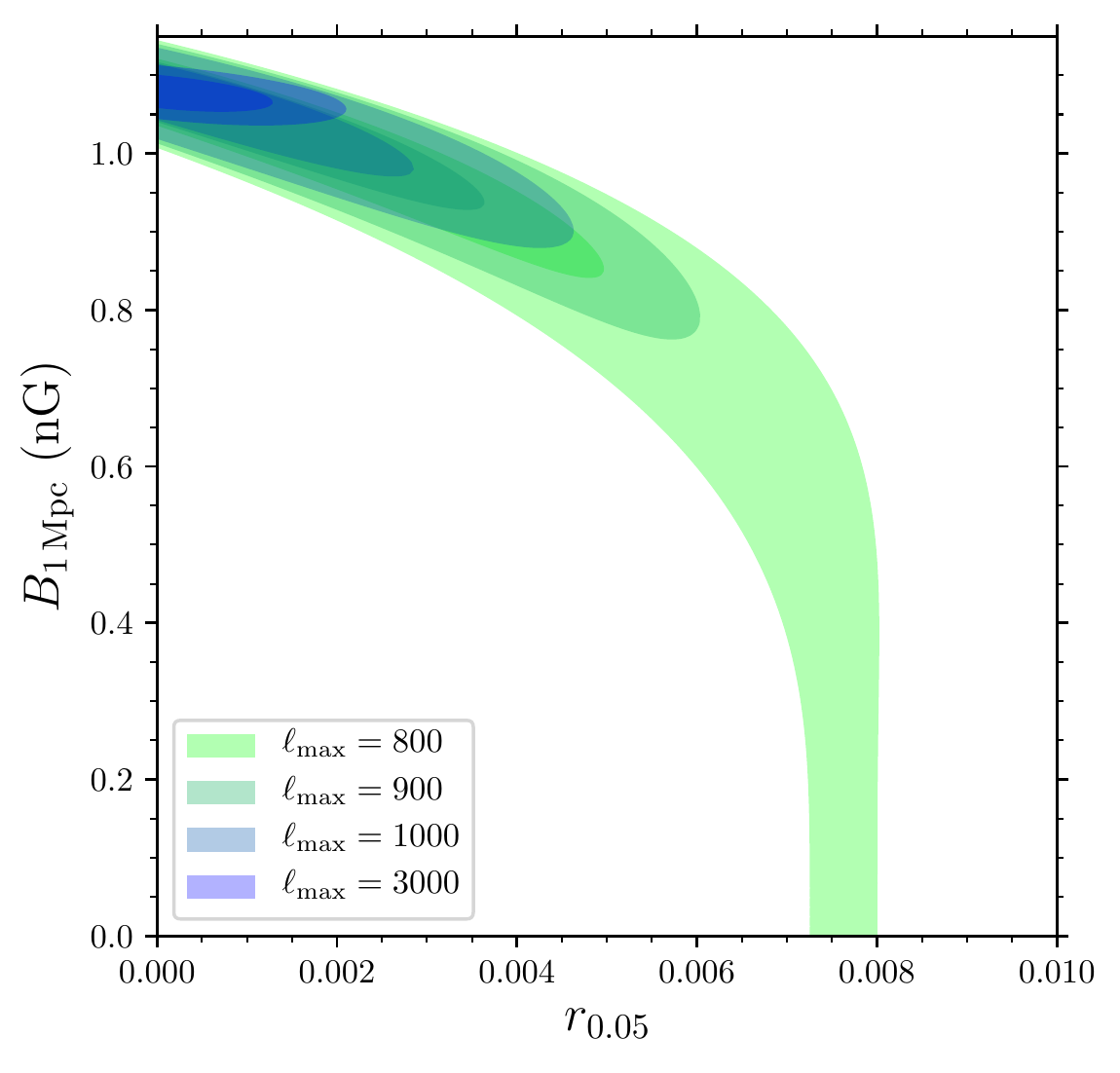}
\caption{Impact of varying $\ell_{\rm max}$ on CMB-S4 constraints: 
we see that the degeneracy between $r$ and $\Bmpc$ is broken if $B$-mode anisotropies can be measured on scales $\ell \gtrsim 900$.}
\label{fig:ell_max}
\end{figure*}

We can show the importance of including small angular scales in a different way 
by using directly the full $B$-mode likelihood for the CMB-S4 experiment, as we did in our analysis of the impact of delensing. 
Varying $\ell_{\rm max}$ (having fixed the delensing parameter $\alpha$ to $1$) 
we can see at which angular scale the degeneracy between $r$ and $\Bmpc$ can be broken by this experiment. 
In Fig.~\ref{fig:ell_max}, we see that this happens at $\ell_{\rm max}\gtrsim900$: 
if CMB-S4 could not access higher multipoles, the constraints on $r$ and $\Bmpc$ would be similar to those of LiteBIRD.

\section{Constraints from Faraday rotation}
\label{sec:5}

\begin{figure*}
\centering
\includegraphics[width=0.7\columnwidth,,keepaspectratio=true]{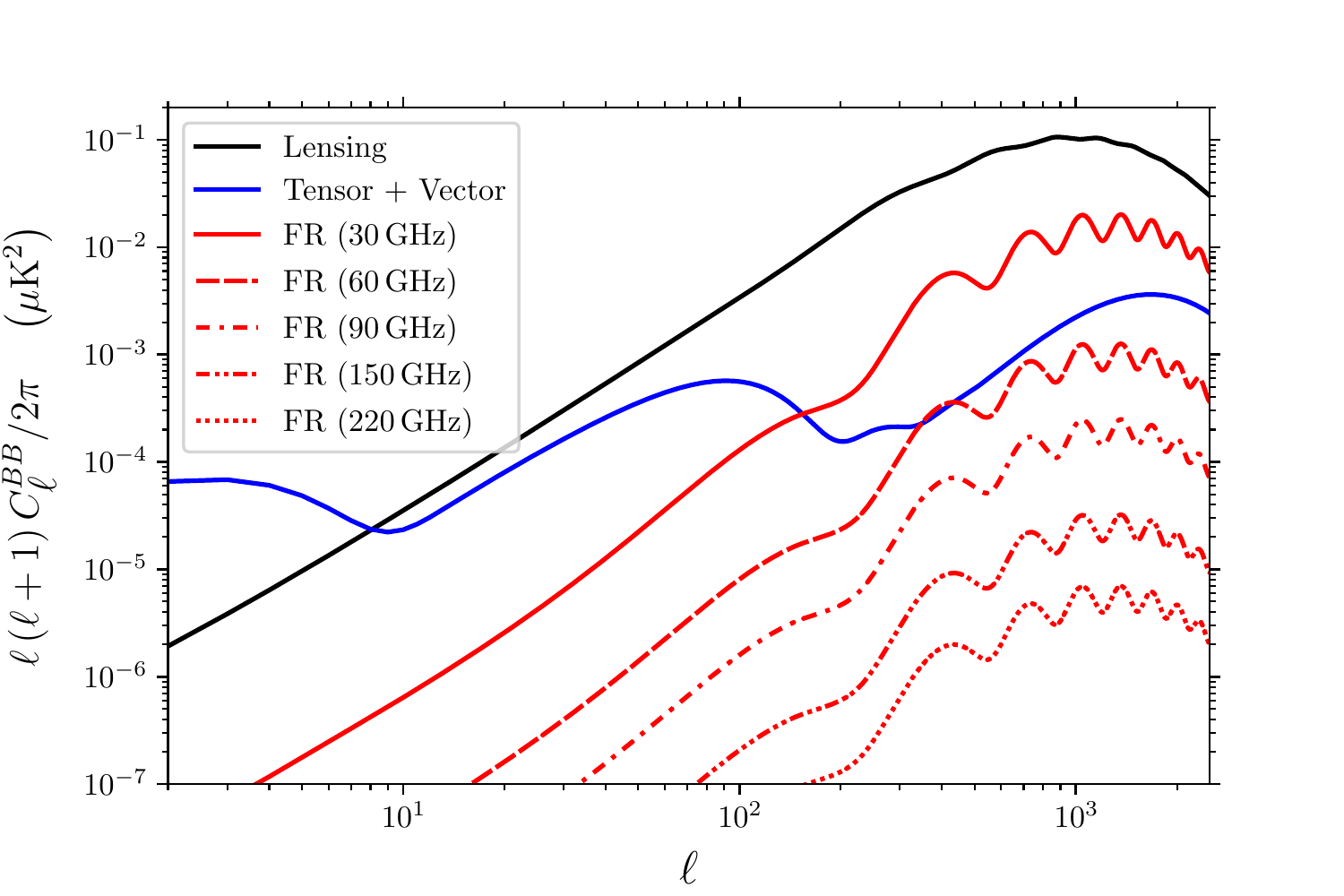}
\caption{$B$-mode angular spectrum from Faraday rotation ($\Bmpc=1.08\,\rm nG$ is assumed) 
at frequencies of $30\,\rm GHz$, $60\,\rm GHz$, $90\,\rm GHz$, $150\,\rm GHz$ and $220\,\rm GHz$. 
The $B$-modes from magnetic perturbations are also shown, together with the standard lensing prediction. 
Clearly the $B$-modes from FR are completely negligible for an experiment mostly sensitive to frequencies around $150\,\rm GHz$ 
as those considered in this paper. PIXIE will operate at lower frequencies, but it will have access only to very low multipoles, 
where the FR signal is negligible anyway.}
\label{fig:FRBB}
\end{figure*}

\noindent PMFs also induce Faraday rotation (FR) of CMB polarization. It is therefore useful to evaluate the ability of future CMB experiments 
to detect a PMF with an amplitude of $1.08\,\mathrm{nG}$ through FR and break the possible degeneracies between $r$ and $\Bmpc$. 

FR of the linear polarization of CMB photons is described by the rotation angle $\alpha_F$, 
defined in terms of the Stokes $Q$ and $U$ parameters by $(Q\pm iU)(\hat{\vec{n}})\to(Q\pm iU)(\hat{\vec{n}})e^{\pm 2i\alpha(\hat{\vec{n}})}$. 
An inhomogeneous magnetic field sources anisotropies in the rotation angle, 
whose angular power spectrum is related to the two-point correlation function of the magnetic field 
(see, e.g., \cite{Kosowsky:2004zh,Pogosian:2011qv,De:2013dra,Ade:2015cao,Array:2017rlf}). 
The angular power spectrum $C^{\alpha_F\alpha_F}_\ell$ of the rotation angle $\alpha_F(\hat{\vec{n}})$ 
can be constrained directly by exploiting the fact that, at first order in $\alpha_F$, 
the off-diagonal elements of the two-point correlation functions of $E$- and $B$-modes are proportional to the rotation field \cite{Yadav:2009eb}. 
One can take advantage of this feature to build a quadratic estimator to measure the anisotropic rotation \cite{Gluscevic:2009mm,Caldwell:2011pu,Yadav:2009eb}. 
The rotation of $E$-modes into $B$-modes, moreover, leads to a contribution $C^{BB,{\rm FR}}_\ell$ that should be in principle added to Eq.~\eqref{sum_spectra} 
(see \cite{Giovannini:2017rbc} for an extensive review). 
However the angular spectrum $C_{\ell}^{\alpha_F\alpha_F}$, and then $C^{BB,{\rm FR}}_\ell$, scales as $\nu^{-4}$: 
thus we expect to have a larger signal for these FR-induced $B$-modes at lower frequencies (e.g. $\nu\sim 30\,{\rm GHz}$), 
that will not be optimally sampled by the experiments considered here.
Indeed, these experiments are not conceived for a measurement at these frequencies: 
\begin{itemize}
\item lower frequencies are more contaminated by galactic foregrounds 
and are mostly used to identify and remove them rather than to extract genuine cosmological information;
\item for a given experimental configuration, lower frequencies are limited by diffraction and have smaller angular resolution. 
Therefore, space experiments such as CORE-M5 or LiteBIRD have been designed with the largest number of 
detectors at frequencies around $\sim 150\,{\rm GHz}$ where the minimal foreground contamination is expected.
\end{itemize}
In Fig.~\ref{fig:FRBB} 
we plot the expected signal in the $B$-mode angular power spectrum from Faraday rotation generated by a magnetic field of $1.08\,\rm{nG}$ for various frequencies. 
The Faraday rotation $B$-modes are obtained using Eq.~(38) of \cite{Pogosian:2011qv}. 
In Fig.~\ref{fig:FRBB} we also plot the lensing $B$-mode spectrum and the frequency-independent contribution of Eq.~\eqref{sum_spectra}, 
i.e. the one generated by vector- and
tensor-mode perturbations in the metric, sourced by the stress-energy in the PMF. 
As we can clearly see, the $B$-modes from Faraday rotation at $150\,\rm{GHz}$ are about four orders of magnitude smaller than the lensing signal, i.e. 
they will be undetectable by an experiment operating at those frequencies. 
At $30\,\rm{GHz}$ the signal is much larger, however none of the experiment considered in this paper will sample this frequency with the exception of PIXIE. 
In this case, however, the angular resolution will not be sufficient to detect the $B$-modes from Faraday rotation, since PIXIE can arrive at most at $\ell_{\rm max} = 500$.

\begin{figure*}
\centering
\begin{tabular}{cc}
\includegraphics[width=0.48\columnwidth,keepaspectratio=true]{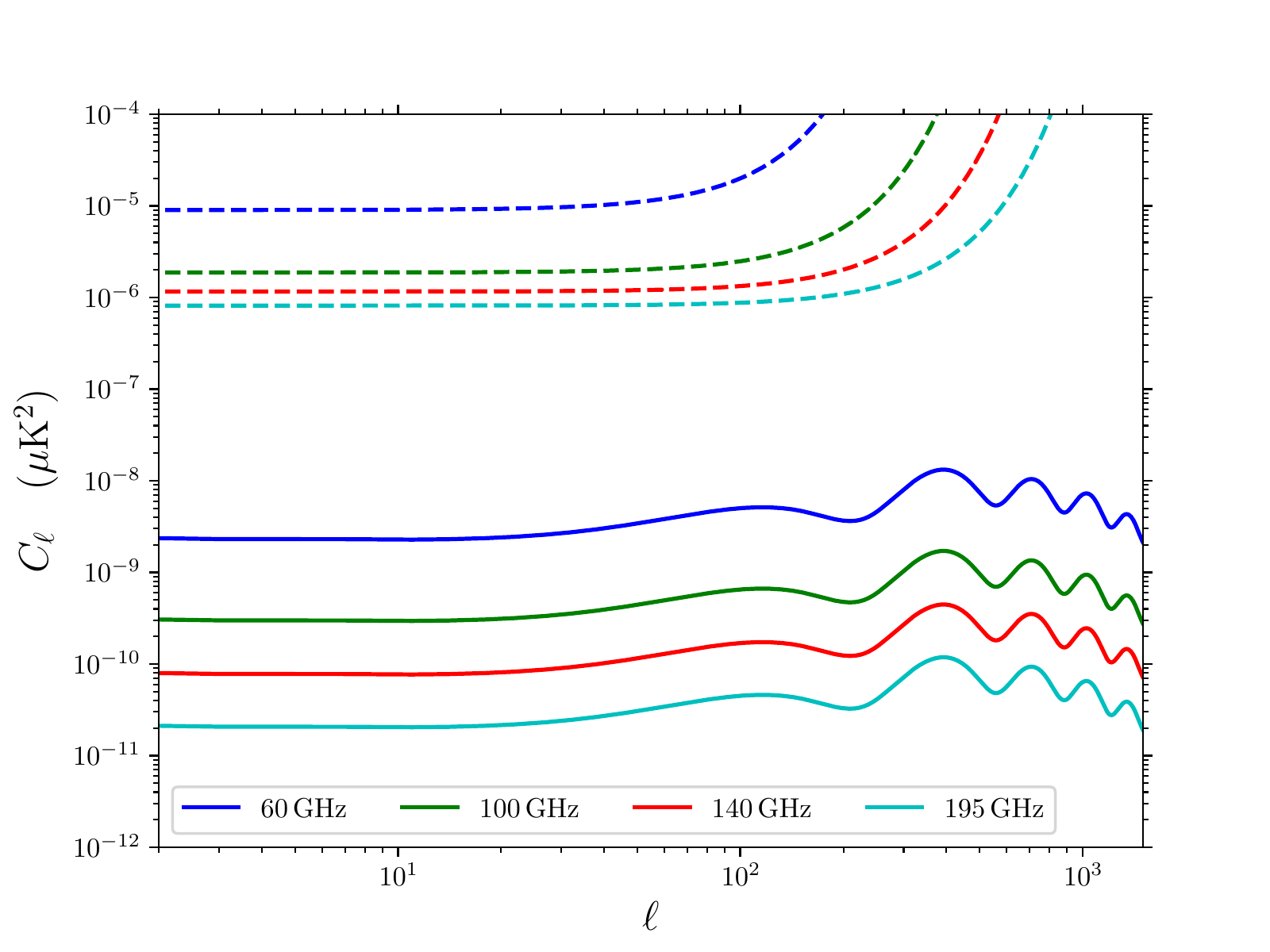} &
\includegraphics[width=0.48\columnwidth,keepaspectratio=true]{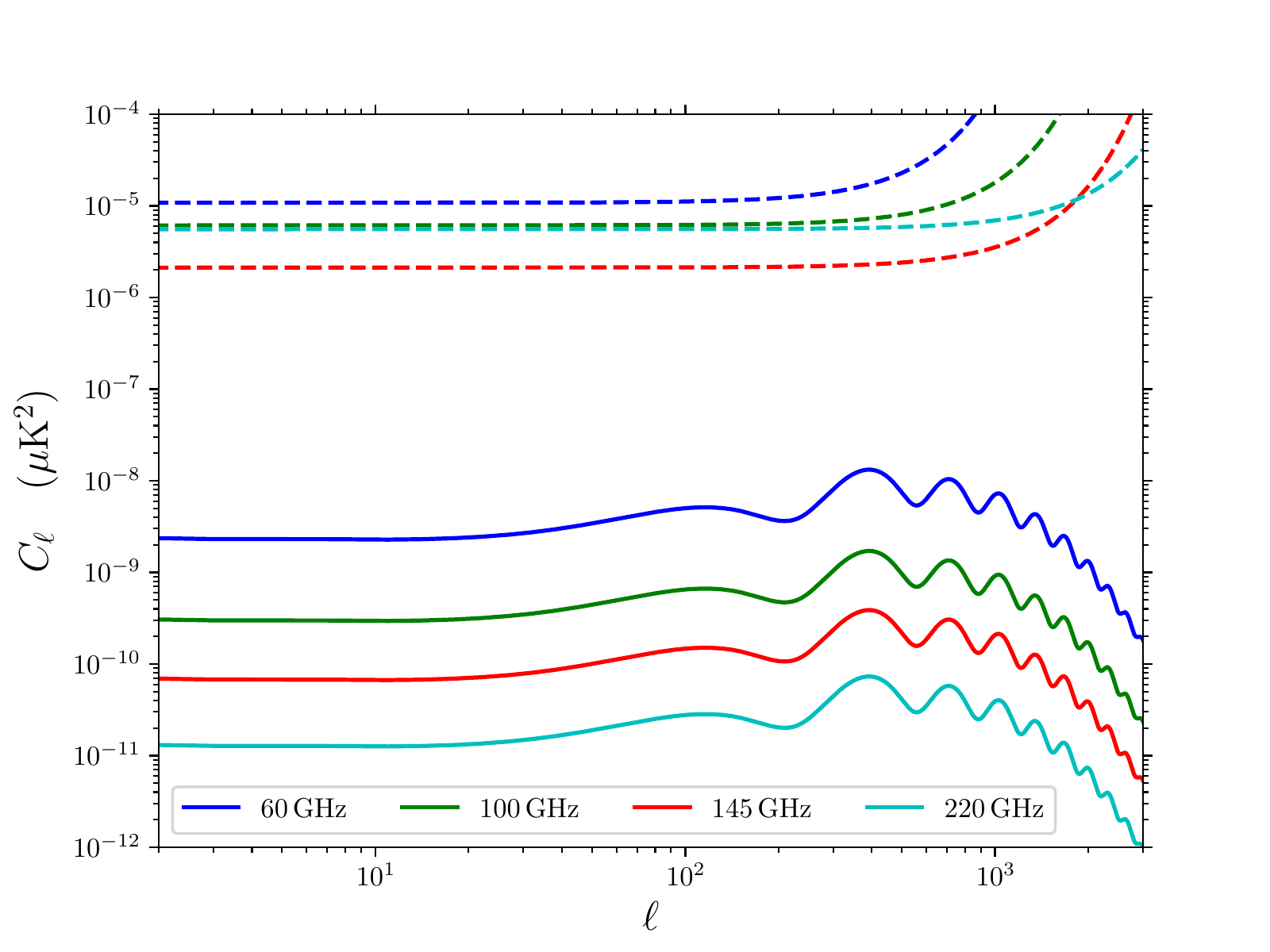}
\end{tabular}
\caption{Expected experimental white noise for the lowest channels of LiteBIRD (top panel) and CORE-M5 (bottom panel), 
together with the expected signal from $B$-modes generated by Faraday rotation (for a PMF with $\Bmpc=1.08\,{\rm nG}$) at the same frequencies. 
Clearly, the signal is undetectable in any frequency channel by any of the two experiments.}
\label{fig:FRexp}
\end{figure*}

The LiteBIRD and CORE-M5 could produce full CMB sky maps at frequencies of $60\,\rm{GHz}$. 
It is therefore interesting to investigate if these $B$-modes from FR could be detected by these experiments. 
Unfortunately, as we show in Fig.~\ref{fig:FRexp}, neither of these two experiments will be able to detect them. 
Indeed, they will not have enough sensitivity and angular resolution at these frequencies.

To summarize, the contribution $C^{BB,{\rm FR}}_\ell$ to the $B$-mode angular power spectrum induced by a PMF of $1.08\,{\rm nG}$ 
is not detectable by the experiments considered here. 
For this reason, in the following we will focus on a forecast for the detection of anisotropies in the FR angle $\alpha_F$ of the $E$ modes: 
clearly, a detection of $\alpha_F$ would allow to confirm whether a possible $B$-mode measurement is due to primordial GWs or to PMFs. 

Then, to perform this forecast we consider the following 
approximated angular power spectrum of $\alpha_F$ 
induced by a nearly scale-invariant PMF of $1.08\,\mathrm{nG}$ (see e.g. Eq. (4) in \cite{Array:2017rlf}):
\begin{equation}
\label{eq:alphaalpha}
\frac{L(L+1)C_{L}^{{\alpha_F}{\alpha_F}}}{2\pi} = 4.2 \times 10^{-8}\,{\rm rad}^2\,\,,
\end{equation}
where we have assumed a CMB observational frequency of $\nu \sim 150\,\mathrm{GHz}$.
We have then estimated the experimental noise on $C_{L}^{\alpha_F \alpha_F}$ 
and computed the relevant signal-to-noise ratio using the quadratic estimator described in 
\cite{Gluscevic:2009mm,Caldwell:2011pu}. 
For simplicity we have neglected the contamination of ``spurious'' FR induced by magnetic fields in our galaxy and lensing. 
Our results (shown in Tab.~\ref{biri}) 
should therefore be considered as optimistic since the removal of these terms could lead to a significantly lower $S/N$ (see \cite{Pogosian:2018vfr}).

\begin{table}
\begin{center}
\begin{tabular}{cccccc}
PIXIE & LiteBIRD & CORE-M5 & Stage-3 (Deep) & Stage-3 (Wide) & CMB-S4 \\
\toprule
\horsp
$0.05$	&$0.7$	&$15$	&$20$	&$2$		&$10^2$
\end{tabular}
\end{center}
\caption{$S/N$ ratio for the detection of a nearly scale-invariant PMF of $1.08\,\mathrm{nG}$ through FR. } 
\label{biri}
\end{table}

We see that experiments as PIXIE and LiteBIRD will essentially be unable to detect our fiducial PMF through FR. 
This is mainly due to the poor angular resolution that does not let these experiments measure $E$- and $B$-modes at $\ell \sim 1000$, 
i.e. at scales that are relevant for a measure of $\alpha_F$. 
Including the frequency dependence of the signal will not change this result since at lower frequencies would correspond also an even lower angular resolution. 
On the contrary, the Stage-3 experiment, especially in the ``deep'' configuration, 
will be able to detect the PMF via FR with high accuracy. 
While in this case $r$ is poorly constrained, a combination with the LiteBIRD experiment could be extremely important.
The CMB-S4 experiment will constrain a PMF with great accuracy, in agreement with the results presented in \cite{Pogosian:2018vfr}.

\section{Discussion and conclusions} 
\label{sec:concl}

\noindent Undoubtedly, one of the main goals of future CMB experiments is a detection of 
inflationary GWs through their effect on the $B$-mode polarization. 
Such a detection would be a strong hint towards the quantum nature of gravity. 
However, a simple detection of $B$-modes is not enough to confirm their primordial origin: 
primordial magnetic fields can cause a contamination of a possible signal from vacuum fluctuations of the metric. 

The goal of this paper was to show that future CMB experiments targeting inflationary GWs at the level of $r\approx\num{d-3}$ 
will not be able to claim a detection unless they are able to distinguish them from a PMF of amplitude $\sim 1\,\mathrm{nG}$. 

Satellite missions as PIXIE or LiteBIRD, that are limited to large angular scales, will not be able to break such degeneracy. 
For experiments with better angular resolution, like CORE-M5 or CMB-S4, it will instead be possible to discriminate between the two mechanisms 
since they will be able to detect the compensated vector perturbations of the PMF. 

A second way to break the degeneracy is that of measuring the PMF through Faraday rotation of the CMB polarization. 
While $B$-modes induced by FR are practically undetectable by the experiments considered here, 
a better opportunity is offered by measurements of anisotropies in the FR angle. 
We find that also in this case PIXIE and LiteBIRD will not be able to significantly detect the PMF. 
However, we have found that the Stage-3 experiment could already put stringent constraints on it. 
A nice complementarity therefore exists between the LiteBIRD and Stage-3 experiments that could allow to break the degeneracy between 
the tensor-to-scalar ratio and the PMF amplitude. 

Before concluding we note that there are other signatures typical of PMFs that could help in distinguishing them from inflationary GWs: 
\begin{itemize}
\item the $B$-modes produced by PMFs are highly non-Gaussian, 
since they are proportional to the square of the field amplitude. 
Consequently, bispectrum and trispectrum measurements could also place strong constraints on them \cite{Ade:2015cva,Hortua:2015jkz,Trivedi:2013wqa};
\item PMFs are damped on small scales, leading to heating of baryons and electrons and producing Compton-$y$ distortions in the CMB (see \cite{Ade:2015cva}). 
This has been used to place an upper limit of $0.90\,\rm nG$ at $95 \%$ C.L on the magnetic field amplitude 
(recently \cite{Paoletti:2018uic} has improved this bound to $0.83\,\rm nG$). 
\end{itemize}
Regarding the second of these signatures, 
we emphasize that uncertainties in the modeling of the heating and in the reionization process may affect the constraint \cite{Chluba:2015lpa}. 
Moreover, and most importantly, 
a PMF of $\sim 0.9\,\rm nG$ would produce a $B$-mode spectrum essentially rescaled by a factor of $0.5$ with respect to the template considered here. 
This could still bias future CMB polarization searches for primordial GWs at the level of $r\approx\num{d-3}$ if not accounted for. 

In summary, future constraints at the level of $0.2\,\mathrm{nG}$, as expected from FR measurements by the CMB-S4 experiment (see also \cite{Pogosian:2018vfr}), 
will be crucial in limiting a spurious $B$-mode PMF contribution to sub-percent level respect to the value of $r\approx\num{d-3}$.

\acknowledgments 

\noindent We are clearly indebted to Levon Pogosian and Alex Zucca for providing their MagCAMB code and for many useful comments and discussions. 
We thank Antony Lewis for providing his CosmoMC code. It is also a pleasure to thank Robert Caldwell, 
Ruth Durrer, Franz Elsner, Takeshi Kobayashi and Eiichiro Komatsu for extremely useful comments and suggestions. 
GC acknowledges support from the Starting Grant (ERC-2015-STG 678652) ``GrInflaGal'' of the European Research Council. 
EDV acknowledges support from the European Research Council in the form of a Consolidator Grant with number 681431.



\bibliographystyle{utphys}
\bibliography{refs}

\end{document}